\documentclass[preprint,review,11pt,authoryear]{elsarticle}

\let\today\relax
\makeatletter
\def\ps@pprintTitle{%
    \let\@oddhead\@empty
    \let\@evenhead\@empty
    \def\@oddfoot{\footnotesize\itshape
         {Preprint submitted to arXiv} \hfill\today}%
    \let\@evenfoot\@oddfoot
    }
\makeatother

\usepackage{amsmath, amsthm}   
\usepackage{booktabs}  
\usepackage{bbm}       
\usepackage{bm}        
\usepackage[lined,boxed,commentsnumbered, ruled]{algorithm2e}   
\usepackage{indentfirst}  
\usepackage{subcaption} 
\usepackage[font=footnotesize]{caption}
\usepackage{enumitem}  
\usepackage[table]{xcolor} 

\usepackage[left=1in,right=1in,top=1in,bottom=1in]{geometry}

\usepackage{pgf}
\usepackage{graphics, color, amsfonts, dsfont, graphicx, amssymb, epsfig}
\usepackage[utf8]{inputenc}  
\usepackage{tikz,pgfplots}
\usepackage[normalem]{ulem}

\usetikzlibrary{matrix,positioning,decorations.pathreplacing}
\usepackage{arydshln}  

\RequirePackage[OT1]{fontenc}
\RequirePackage{natbib}   
\RequirePackage[colorlinks,citecolor=blue,urlcolor=blue]{hyperref}

\graphicspath{ {Figures/} }

\newtheorem{thm}{Theorem}
\newtheorem{lem}[thm]{Lemma}
\newtheorem{prop}[thm]{Proposition}
\newtheorem{cor}[thm]{Corollary}

\theoremstyle{definition}

\theoremstyle{remark}
\newtheorem{rem}{\textbf{Remark}}

\newcommand*{\appheading}[1][Appendix]{%
  \setcounter{secnumdepth}{0}\section{#1}\setcounter{secnumdepth}{3}%
  \numberwithin{table}{section}
  \numberwithin{figure}{section}
}

\DeclareMathOperator*{\argmax}{arg\,max}

\DeclareMathOperator{\diag}{diag}

\newcommand{\ra}[1]{\renewcommand{\arraystretch}{#1}}  

\newcommand{\R}{\mathbb{R}}

\newcommand{\norm}[1]{\left\lVert#1\right\rVert}
\newcommand{\norms}[1]{\lVert#1\rVert}

\newcommand{\calI}{\mathcal{I}}

\newcommand{\calU}{\mathcal{U}}

\newcommand{\one}{\bm{1}}
\newcommand{\zero}{\bm{0}}
\newcommand{\tp}{\top}

\newcommand{\xt}{\tilde{x}}

\newcommand{\bb}{\textup{\textbf{b}}}
\newcommand{\bh}{\hat{b}}
\newcommand{\pb}{\textbf{p}}
\newcommand{\ub}{\textup{\textbf{u}}}
\newcommand{\vb}{\textup{\textbf{v}}}

\newcommand{\xb}{\textup{\textbf{x}}}
\newcommand{\yb}{\textup{\textbf{y}}}
\newcommand{\ybt}{\tilde{\yb}}

\newcommand{\Cb}{\textup{\textbf{C}}}

\newcommand{\Ub}{\textbf{\textup{U}}}

\newcommand{\xbt}{\tilde{\xb}}

\newcommand{\bbh}{\hat{\bb}}

\newcommand{\wt}{\tilde{w}}
\newcommand{\bbt}{\tilde{\bb}}
\newcommand{\wbar}{\bar{w}}
\newcommand{\bbbar}{\bar{\bb}}

\newcommand{\Thetab}{\bm{\Theta}}
\newcommand{\Sigmab}{\bm{\Sigma}}

\newenvironment{framework}[1][htb]
  {
  \begin{algorithm}[#1]%
  }{\end{algorithm}}

\begin{document}

\begin{frontmatter}

\title{Adaptive Robust Online Portfolio Selection}

\author[mymainaddress1]{Man Yiu Tsang}
\cortext[cor1]{Corresponding author. }
 \ead{mat420@lehigh.edu}
\author[mymainaddress2]{Tony Sit\corref{cor1}}
 \ead{tony.sit@cuhk.edu.hk}
\author[mymainaddress2]{Hoi Ying Wong}
 \ead{hywong@cuhk.edu.hk}

 \address[mymainaddress1]{Department of Industrial and Systems Engineering, Lehigh University, Bethlehem, PA,  USA \vspace{2mm}}
 \address[mymainaddress2]{Department of Statistics, The Chinese University of Hong Kong, Shatin, N.T., Hong Kong}

\begin{abstract}

\noindent  The online portfolio selection (OLPS) problem differs from classical portfolio model problems, as it involves making sequential investment decisions. Many OLPS strategies described in the literature capture market movement based on various beliefs and are shown to be profitable. In this paper, we propose a robust optimization (RO)-based strategy that takes transaction costs into account. Moreover, unlike existing studies that calibrate model parameters from benchmark data sets, we develop a novel adaptive scheme that decides the parameters sequentially. With a wide range of parameters as input, our scheme captures market uptrend and protects against market downtrend while controlling trading frequency to avoid excessive transaction costs. We numerically demonstrate the advantages of our adaptive scheme against several benchmarks under various settings. Our adaptive scheme may also be useful  in general sequential decision-making problems. Finally, we compare the performance of our strategy with that of existing OLPS strategies using both benchmark and newly collected data sets. Our strategy outperforms these existing OLPS strategies in terms of cumulative returns and competitive Sharpe ratios across diversified data sets, demonstrating its adaptability-driven superiority.


\begin{keyword} 
Online portfolio selection, robust optimization, adaptive parameter selection, transaction costs
\end{keyword}

\end{abstract}
\end{frontmatter}


%




\section{Introduction} \label{sec:introduction}
Portfolio optimization problems have been extensively studied since the seminal work of \cite{Markowitz:1952} on mean-variance portfolio theory. These problems concern investment decisions that an investor has to make when facing uncertain returns from various assets. Consequently, portfolio models often incorporate a trade-off between return and risk. Recent literature develops various types of portfolio optimization models, such as multi-period or continuous-time portfolio models \citep{Li_Ng:2000,Zhou_Li:2000} or those that optimize the portfolio constructed with respect to different types of risk measures \citep{Ahmadi-Javid_Fallah-Tafti:2019,Rockafellar_Uryasev:2000,Sereda_et_al:2010}. 

Unlike classical portfolio optimization model problems, which are usually based on distributional information of historical returns, the online portfolio selection (OLPS) problem requires investment decisions to be made sequentially based on recent return information \citep{Cover:1991}. Thus, a distinctive feature of the OLPS problem is the prediction of the next period return based on certain financial beliefs or machine learning tools \citep{Das_et_al:2013, Lai_et_al:2018, Li_et_al:2012, Li_et_al:2015}. This feature distinguishes the OLPS problem or the short-term portfolio optimization problem from the long-term portfolio optimization problem. The OLPS problem is typically solved to formulate OLPS strategies that maximize the cumulative wealth over a specific investment horizon. Thus, in each period, investors must decide on the composition of their new portfolio based on all of the historical data. This requires sequential decisions to be made (e.g., daily rebalancing), meaning that tractable and computationally efficient OLPS strategies are generally preferred. 

Robust optimization is now more prevalent in the literature than in previous years. This reflects the fact that while solving the (risk-neutral) stochastic programming problem may often generate optimistic solutions, robust optimization mitigates the negative effects of unfavorable scenarios and model misspecification (see, for instance, \citealp{Ben-Tal_et_al:2009}). This is particularly useful when the underlying distribution of randomness is unknown as the stochastic programming approach requires knowledge of the distribution that this randomness follows. Furthermore, due to the development of robust optimization theories (see, for instance, \citealp{Bertsimas_et_al:2011}), robust optimization is widely adopted in various practical research areas and applications, such as those related to supply chain or inventory \citep{Bertsimas_Thiele:2006, Kim_et_al:2018, Solyali_et_al:2016}, energy \citep{Louca_Bitar:2018,Zhang_et_al:2018,Zhou_et_al:2018} or health care \citep{Rath_et_al:2017,Shi_et_al:2019,Unkelbach_Paganetti:2018}.

The consideration of robustness in the portfolio optimization model has become more popular in recent years \citep{Fabozzi_et_al:2007, Huang_et_al:2010}. This is because with an unknown distribution of random asset returns, risk-averse investors may decline to use the empirical distribution (based on historical returns) but instead prefer to devise a strategy that can hedge against worst-case scenarios according to their risk preferences. In other words, they may seek a robust portfolio that can mitigate losses they may incur from unfavorable scenarios. For example, if an actual distribution behaves differently from the empirical distribution adopted  (e.g., a sudden market downturn occurs), a robust portfolio may exhibit better out-of-sample performance than other portfolios \citep{Santos:2010}. However, to the best of our knowledge, the use of robustness in OLPS strategies remains rare, despite the fact that it can assist investors to make decisions based on noisy data and limited distributional information \citep{Huang_et_al:2016, Lai_et_al:2020}.

\begin{figure}
    \centering
    \includegraphics[scale=0.70]{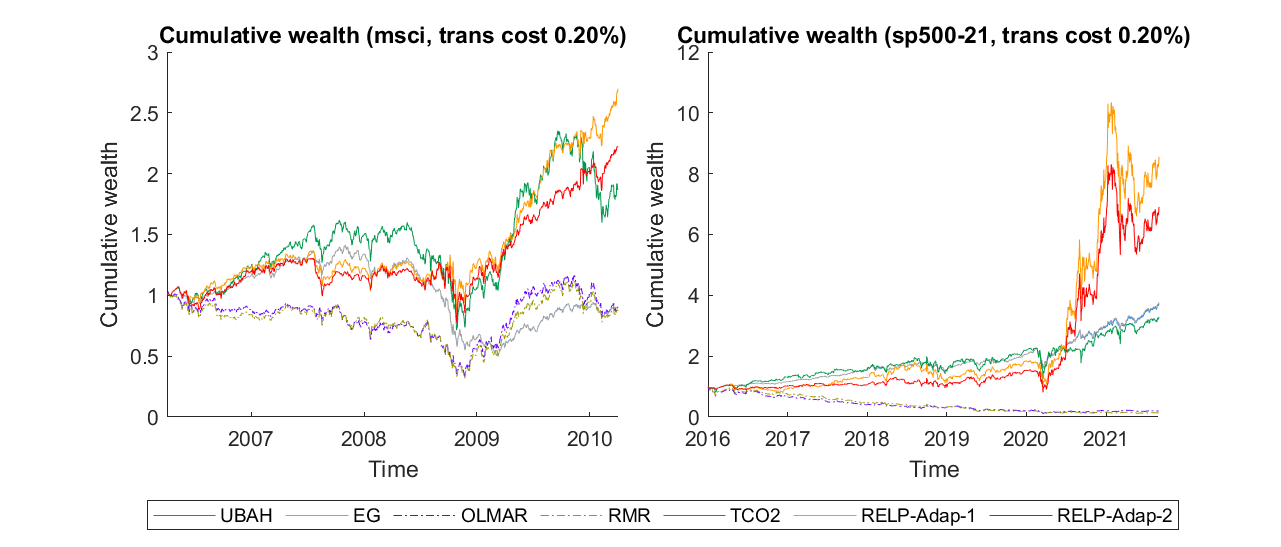}
    \caption{Cumulative wealth generated by various OLPS strategies applied to the MSCI and SP500-21 data sets and with a $0.2\%$ transaction cost. Transaction costs (or regularization) are explicitly considered by the EG, TCO2, RELP-Adap-1 and RELP-Adap2 strategies.}
    \label{fig:demo}
\end{figure}

The transaction costs associated with the OLPS strategy are also critical, as the trading volume may be tremendous in the OLPS problem (e.g., due to daily rebalancing). Therefore, there is a trade-off between active rebalancing and transaction costs. Various models have been proposed in literature to specifically manage transaction costs. For instance, some models enforce an $\ell_1$ norm penalty on a portfolio change to control trading frequency \citep{Li_et_al:2018, Guo_et_al:2021}. In Figure~\ref{fig:demo}, active OLPS strategies that do not consider transaction costs (e.g., OLMAR and RMR) suffer from over rebalancing, which leads to poor portfolio performance. Indeed, their average turnovers (ATs)---measured as the normalized change in two consecutive portfolio proportions (defined in \eqref{eqn:average_turnover})---are greater than $60\%$, which is significantly different from zero. In contrast, the ATs of the strategies we develop (RELP-Adap-1 and RELP-Adap-2) range only from $5\%$ to $16\%$, implying that they result in a less frequent change in portfolio weights than the said  strategies.


Despite the fact that it is critical to obtain a suitable choice of parameters a priori in an OLPS strategy, where these parameters are relevant to the performance of the strategy, the existing literature tends to calibrate these parameters empirically. However, such default parameters suggested in the literature may render those strategies futile in practice. For example, Figure~\ref{fig:demo} reveals that, with the default choice of parameters, TCO2 performs reasonably well when applied to the benchmark MSCI data set but fails to capture huge profits  when applied to our newly collected SP500-21 data set (see \ref{appdx:new_data_set} for details). In contrast, the adaptive parameter scheme in the strategies we develop protects against market downturn (the 2008 financial crisis in the MSCI data set) and captures rapid market upturn (in 2020 in the SP500-21 data set). 

%

The contributions of this paper are as follows.

\begin{itemize}
\item We develop a novel OLPS strategy that incorporates robustness and the consideration of transaction costs. Specifically, we employ an ellipsoidal uncertainty set that exploits the sample variance (i.e., second order information) of a return as its shape parameter. We then reformulate our novel OLPS strategy as a second-order conic program (SOCP), which can readily be solved using off-the-shelf optimization solvers.

\item We develop an adaptive scheme that relies only on a (possibly wide) range of values of parameters as input. In particular, we borrow the idea of the classical selection of the best problem (see, for instance, \citealp{Branke_et_al:2007,Fan_et_al:2016,Kim_Nelson:2006}). We demonstrate the advantages of our novel adaptive scheme by comparing its  performance with those of some benchmark adaptive schemes. Our novel adaptive scheme is designed to maximize the cumulative wealth of expert portfolios that use different parameters (corresponding to differing extents of tradeoff between risk and return). This assists in both capturing market upturn and protecting against market downturn, and thus yields sufficient cumulative wealth to match the risk appetites of the investors concerned.

\item We use our novel model with an adaptive scheme to perform numerical experiments on benchmark data sets and three new testing data sets. Our strategy yields the three greatest amounts of cumulative wealth in $13$ out of $18$ settings (comprising various data sets and transaction costs). In particular, in nearly $40\%$ of the settings, the cumulative wealth obtained using our strategy is at least $50\%$ greater than that obtained using the other tested strategies. These results suggest that the benchmark data sets widely adopted in the OLPS literature \citep{Li_Hoi:2018} may be biased toward certain sets of parameters. This highlights the importance of adaptive parameter selection, which is seldom investigated in the literature.
\end{itemize}

The remainder of this paper is organized as follows. In Section~\ref{sec:literature_review}, we review related literature on OLPS strategies. In Section~\ref{sec:robust_model}, we detail the development of our model and derive its reformulation, while in Section~\ref{sec:adaptive_scheme} we discuss our adaptive scheme in terms of the selection of parameters in the model, based on the selection of the best idea. In Section~\ref{sec:numerics}, we perform numerical experiments to examine the performance of our model and adaptive scheme. We conclude the paper in Section~\ref{sec:conclusion}, and we relegate all of the proofs to \ref{appdx:proofs}.


\section{Literature review} \label{sec:literature_review}
In this section, we briefly review classes of OLPS strategies described in the literature. For a detailed recent survey, we refer readers to \cite{Li_Hoi:2014}. Two benchmark strategies can be used to examine the performance of an OLPS strategy \citep{Li_Hoi:2014}: the uniform buy-and-hold (UBAH) strategy and the uniform constant rebalanced portfolio (UCRP) strategy. The UBAH strategy, which is also known as the market strategy, begins with a portfolio with equal weights on each asset, without any rebalancing during the investment horizon. The UCRP strategy rebalances a portfolio to an equally weighted portfolio at the end of a period, as the portfolio proportion changes after accounting for the return of the previous period. 

OLPS strategies are grouped into several classes \citep{Li_Hoi:2014}, two of which are discussed here. The first class are return momentum or ``follow the winner'' strategies, which take advantage of the best-performing assets/portfolios in a set of portfolios. For instance, the current return vector may be used as the return predictor. Some return momentum strategies may also use a regularization term on portfolio weights \citep{Agarwal_et_al:2006, Cover:1991, Helmbold_et_al:1998, Lai_et_al:2017}. The second class are mean reversion belief-based strategies, which involve the purchase of assets that have performed poorly in the previous period and the selling of those that have performed well. Mean reversion behaviour can be captured by asset correlations, a return threshold (that dictates whether trading should be active or passive) or the prediction of mean-reverting returns \citep{Borodin_et_al:2004, Li_et_al:2011b, Li_et_al:2012, Li_et_al:2013, Li_et_al:2015, Guo_et_al:2021}. Aside from these two classes of strategies, patching-matching-based strategies are also studied; these aim to construct a return predictor by searching for similar patterns in historical returns. Several similarity measures are described in the literature, such as Euclidean distance, $k$-nearest neighborhood and correlation based measures \citep{Gyorfi_et_al:2006, Gyorfi_et_al:2008, Li_et_al:2011}. We refer readers to \ref{appdx:OLPS_strategy} for a more detailed description of these strategies.

However, few studies consider the transaction costs incurred in a portfolio model, despite this being a critical consideration for a frequently rebalanced portfolio. In one example, \cite{Li_et_al:2018} develop a realistic framework for the OLPS problem with consideration of transaction costs, which we adopt in this paper. Moreover, \cite{Li_et_al:2018} suggest the use of an $\ell_1$ penalty in the objective function, known as the TCO strategies, to account for the costs incurred during a transaction. Their experiments show that TCO strategies perform better than previous strategies in the presence of transaction costs. \cite{Guo_et_al:2021} address the issue of transaction costs in their optimization model in a similar fashion, by reformulating the model into a linear program (LP). \cite{Shen_et_al:2014} develop the doubly regularized portfolio model, which employs an $\ell_1$ penalty on the portfolio weight and an $\ell_2$ penalty on the change in the portfolio weight. Furthermore, \cite{Lai_et_al:2018} develop a sparse portfolio model and a converging alternative direction method of multipliers algorithm to solve.

Similarly, the use of robustness in OLPS models is not extensively studied in the literature. \cite{Huang_et_al:2016} exploit a robust estimate of the unknown return by performing minimization using an $\ell_1$ penalty and the Huber loss function to obtain the price estimate; this is known as the RMR strategy. \cite{Lai_et_al:2020} consider a mean-variance model for the OLPS problem by developing a rank-one covariance matrix estimate. In their portfolio model, the worst-case return over historical returns within a certain-sized window is used in the objective function. To the best of our knowledge, no other studies apply robust portfolio models to solve the OLPS problem. Thus, herein, we develop a novel OLPS strategy that takes into account both transaction costs and robustness. This distinguishes our efforts from those of previous researchers.

\section{A robust online portfolio strategy with consideration of transaction costs} \label{sec:robust_model}
\subsection{Problem setting}  \label{subsec:problem}
In the OLPS problem, we assume that there are $m$ assets in the market and that we want to invest in $n$ consecutive periods. The (closing) price of asset $i$ at period $t$ is $p_{t,i}$, and we use $\pb_t=(p_{t,1},\dots,p_{t,m})^\tp$ as the price vector. The change in price is captured by the price relative vector $\xb_t=\pb_t/\pb_{t-1}$, with vector division performed elementwise. In each period, we search for a portfolio $\bb_t\in\Delta_m$, where $\Delta_m=\{\bb\in\R^m\mid \bb \geq \zero,\, \bb^\tp\one = 1\}$. Here, $\bb$ denotes the proportion of wealth we want to invest in each asset. The simplex set $\Delta_m$ is used to ensure that our strategy is self-financed without the need for short selling. The initial portfolio is set to $\bb_1=(1/m,\dots,1/m)^\tp$. After implementing the decision $\bb_t$ at period $t$, the price relative vector $\xb_t$ is observed and the end-of-period portfolio is 
$$\bbh_t = \frac{\bb_t\cdot \xb_t}{\bb_t^\tp \xb_t},$$
where $\cdot$ denotes elementwise multiplication. Note that the ratio $\bbh_t$ is different from $\bb_t$, due to the random return $\xb_t$. 

To accommodate the transaction cost incurred during portfolio rebalancing, we adopt the framework that \cite{Li_et_al:2018} develop for the OLPS problem. We impose a proportional transaction cost if we trade the assets. Following the literature on the OLPS problem, we assume that the proportional transaction cost $\gamma<1$ is the same for purchasing and selling the assets \citep{Gyorfi_Vajda:2008, Li_et_al:2018}. For a given portfolio $\bb_{t+1}$, let $w_t$ be the net proportion remaining after accounting for the transaction cost . Then, $w_t$ is the solution to the following equation
\begin{equation} \label{eqn:weight_balance}
w_t+\gamma\Bigg[\sum_{i=1}^m \Big(\bh_{t,i}-w_t b_{t+1,i}\Big)^+ + \sum_{i=1}^m \Big(w_t b_{t+1,i}-\bh_{t,i}\Big)^+ \Bigg]=1,
\end{equation}
where $(a)^+=\max\{a,0\}$. This is equivalent to
$$w_t + \gamma\norms{\bbh_t- w_t\bb_{t+1}}_1 = 1.$$
The two terms inside the square brackets in \eqref{eqn:weight_balance} correspond to the selling and purchasing proportions, respectively. For instance, if rebalancing requires 2\% of the wealth to be used to cover transaction costs, the net weight $w$ is simply $0.98$. Let $S_t$ be the cumulative wealth at the end of period $t$. Then, the cumulative wealth is updated to $S_{t+1}=S_t w_t (\bb_{t+1}^\tp \xb_{t+1})$. Framework~\ref{framework:OLPS_TC} details the entire investment process. 

\LinesNumbered  
\IncMargin{1em}
\begin{framework}[t]  
\SetKwInOut{Initialization}{Initialization}

\Initialization{Set $\bb_1=\one/m$, $\bb_0=\zero$ and $S_0=1$.} 
\For{$t=1,\dots,n$}{
Obtain the new portfolio weights $\bb_t$ based on an OLPS strategy.  \\
Compute the net proportion $w_{t-1}$ of the rebalancing (from $\bbh_{t-1}$) by solving \eqref{eqn:weight_balance}.  \\
Observe the realized return $\xb_t$.  \\
Update the cumulative wealth as $S_t = S_{t-1} w_{t-1}(\bb_{t}^\tp \xb_{t})$.  \\
Obtain the current portfolio weights $\bbh_t = \bb_t\cdot \xb_t / \bb_t^\tp \xb_t$. 
}
\BlankLine
\caption{A framework for the OLPS problem with consideration of transaction costs}\label{framework:OLPS_TC}
\end{framework}\DecMargin{1em}

\subsection{Base model} \label{subsec:base_model}
OLPS strategies often incorporate a prediction step, i.e., the prediction of $\xb_t$, and then use the predicted value $\xbt_t$ in an optimization step. For instance, the multi-period moving average reversion can be used, and is given by
\begin{equation} \label{eqn:MAR_predictor}
\xbt_t=\frac{(1/W)\sum_{i=t-W+1}^t \pb_i }{\pb_t}= \frac{1}{W}\left(1+\frac{1}{\xb_{t}}+\cdots+\frac{1}{\bigodot_{i=0}^{W-2} \xb_{t-i}}\right)
\end{equation}
as discussed in \cite{Li_et_al:2015}, where $W$ is the moving average window size and $\bigodot$ denotes the elementwise product. We also adopt this predictor in our numerical experiments. We emphasize that the model we develop and our adaptive scheme are not restricted to the particular choice of $\xbt_t$ specified in \eqref{eqn:MAR_predictor}; they can incorporate any favourable predictor $\xbt_t$. In the base model, we consider the arithmetic return objective, i.e., $\bb^\tp \xb_t$, which is the goal of the OLPS problem. Simultaneously, we control the transaction costs when we rebalance $\bbh_{t-1}$ to $\bb_t$. That is, we penalize excessively frequent trading. Proposition 4.1 from \cite{Li_et_al:2018} shows that the net proportion $w_{t-1}$ satisfies
$$\frac{1-\gamma}{1-\gamma+\gamma\norms{\bbh_{t-1}-\bb_t}_1} \leq w_{t-1} \leq \frac{1+\gamma}{1+\gamma+\gamma\norms{\bbh_{t-1}-\bb_t}_1}.$$
Given this, \cite{Li_et_al:2018} suggest the use of an $\ell_1$ penalty (i.e., $\lambda\norms{\bbh_{t-1}-\bb_t}_1$) in the objective function to control the transaction costs. However, no adjustments for transaction costs are made in the return component in the objective function, which does not reflect real practice; i.e., in reality, investors are more concerned about the magnitude of a transaction cost-adjusted return than about the magnitude of the return itself \citep{Arnott_Wagner:1990}. Moreover, the goal of the OLPS problem is to maximize cumulative wealth, which is the product of transaction cost-adjusted period returns. Therefore, in our base model, we explicitly consider the transaction cost-adjusted return (see, for instance, \citealp{Lobo_et_al:2007}). Our base model is as follows:
\begin{equation} \label{eqn:TC}
\begin{aligned}
& \underset{w\geq 0,\,\bb\geq \zero}{\text{maximize}}
& & w\bb^\tp \xbt_t-\lambda\norms{\bbh_{t-1}-w\bb}_1\\
& \text{subject to}
& & w + \gamma\norms{\bbh_{t-1}-w\bb}_1=1,\quad \bb^\tp \one =1,  \\
\end{aligned}
\end{equation}
where the parameter $\lambda>0$ serves as a rebalancing penalty and $w\bb$ is the transaction cost-adjusted portfolio, which means that $(w\bb)^\tp\one$ can be less than $1$. The first constraint states that the proportions of portfolio values and transaction costs sum to one. Optimization of the first term in the objective function is our primary goal, i.e., maximization of the transaction cost-adjusted return. The second term controls the trading frequency or volume in the form of a penalty.

Problem \eqref{eqn:TC} is a potentially difficult optimization problem for two reasons. First, the quadratic term $w\bb$ means that the formulation is a non-linear optimization problem. Second, the first constraint associated with the balance equation is non-convex due to the $\ell_1$ equality constraint. Hence, problem \eqref{eqn:TC} is a non-convex program with quadratic variable terms. Nevertheless, we show that the problem is equivalent to a convex program, and can ultimately be reformulated as an LP. Unlike in the classical setting, wherein the return is maximized, we use an extra non-linear term due to the $\ell_1$ penalty. This complicates the reformulation of the model.

\begin{thm} \label{thm:TC_convex}
Consider the following convex program
\begin{equation} \label{eqn:TC_convex}
\begin{aligned}
& \underset{\bb\geq\zero}{\textup{maximize}}
& & \bb^\tp \xbt_t-\lambda\norms{\bbh_{t-1}-\bb}_1\\
& \textup{subject to}
& & \bb^\tp\one + \gamma\norms{\bbh_{t-1}-\bb}_1 \leq 1,\\
\end{aligned}
\end{equation}
and let $\bb_{(2)}^*$ be an optimal solution. Then, the optimal value of \eqref{eqn:TC} is the same as that of \eqref{eqn:TC_convex}, and $(w_{(1)}^*,\bb_{(1)}^*):=(\one^\tp\bb_{(2)}^* \bb_{(2)}^*/\one^\tp\bb_{(2)}^*)$ is an optimal solution to \eqref{eqn:TC}. 
\end{thm}

\begin{rem}
Note that model \eqref{eqn:TC_convex} remains interpretable after reformulation, and the variable $\bb$ corresponds to the transaction cost-adjusted portfolio. Hence, in general, we have $\bb^\tp\one \leq 1$. Thus, the difference between $1$ and $\bb^\tp\one$ gives the transaction cost proportion $w$.
\end{rem}

\begin{cor} \label{coro:TC_LP}
The problem \eqref{eqn:TC} is equivalent to the following LP.
\begin{equation} \label{eqn:TC_LP}
\begin{aligned}
& \underset{\bb\geq\zero,\,\ub\geq\zero,\,\vb\geq\zero}{\textup{maximize}}
& & \bb^\tp \xbt_t-\lambda(\ub+\vb)^\tp\one\\
& \textup{subject to}
& & \bb^\tp\one + \gamma(\ub+\vb)^\tp\one \leq 1, \\
& & & \ub \geq \bbh_{t-1}-\bb, \quad \vb \geq \bb - \bbh_{t-1}.  \\
\end{aligned}
\end{equation}
\end{cor}

\subsection{Robust model with an ellipsoidal uncertainty set} \label{subsec:robust_model}
Two problems arise if we apply model \eqref{eqn:TC} directly for designing a portfolio. First, due to the linearity of the arithmetic return $\bb^\tp\xbt_t$ in $\bb$, the resulting portfolio will typically consist of investment in a single asset: that with the largest predicted return $\xt_{t,i}$. This is because the standard basis vectors in $\R^m$ are potentially the vertices in the feasibility set. Thus, the portfolio may not be sufficiently diversified and hence suffer from high volatility. Second, the model relies heavily on the prediction of the next-period return $\xbt_t$. However, although portfolios are rebalanced rapidly in solutions to the OLPS problem, this prediction may exhibit a substantial level of uncertainty.

Accordingly, rather than using the arithmetic return itself, and thus possibly obtaining an inaccurate prediction, we resort to the robust optimization approach. The corresponding model is given by
\begin{equation} \label{eqn:TC_robust}
\begin{aligned}
& \underset{w\geq0,\,\bb\geq\zero}{\text{maximize}}
& & \min_{\xb\in\calU} \big\{w\bb^\tp \xb \big\}-\lambda\norms{\bbh_{t-1}-w\bb}_1\\
& \text{subject to}
& & w + \gamma\norms{\bbh_{t-1}-w\bb}_1=1, \quad \bb^\tp \one =1,
\end{aligned}
\end{equation}
where $\calU$ is an uncertainty set. In this paper, we use an ellipsoidal set of the form
\begin{equation} \label{eqn:ellipsoidal_set}
\calU(\xbt_t,\kappa)=\Big\{\xb\in\R^m \, \Big| \,(\xb-\xbt_t)^\tp\Sigmab^{-1}(\xb-\xbt_t) \leq \kappa^2\Big\},
\end{equation}
where $\xbt_t\in\R^m$, $\Sigmab\succ\zero$ and $\kappa\geq 0$ are the center, shape parameter and size of the ellipsoid, respectively \citep{Goldfarb_Iyengar:2003}. In particular, we select $\xbt_t$ as the moving average reversion estimate \eqref{eqn:MAR_predictor} and $\Sigmab$ as the sample covariance matrix, using the most recent $(m+1)$ samples. Employing such a choice of $\Sigmab$ captures the most recent market movement while also satisfying the requirement for positive definiteness. 

It is generally difficult to solve model \eqref{eqn:TC_robust} directly, as it involves an inner maximization problem corresponding to the robustness in $\xb$. Therefore, we must first reformulate this inner maximization problem. The reformulation is detailed in the literature, and for completeness we provide the proof in \ref{appdx:proofs}.

\begin{prop} \label{prop:inner_max}
Given the ellipsoidal uncertainty \eqref{eqn:ellipsoidal_set}, we have
$$\min_{\xb\in\calU(\xbt_t,\kappa)} \big\{\bb^\tp \xb \big\} = \xbt^\tp\bb - \kappa \norms{\Ub\bb}_2,$$
where $\Ub$ is the upper triangular matrix of the Cholesky factorization of $\Sigmab$, i.e., $\Sigmab = \Ub^\tp\Ub$.
\end{prop}

According to Proposition~\ref{prop:inner_max}, the $\ell_2$ norm term corresponds to the standard deviation, i.e.,
$$\norm{\Ub\bb}_2=\sqrt{(\Ub\bb)^\tp(\Ub\bb)}=\sqrt{\bb^\tp\Ub^\tp\Ub\bb}=\sqrt{\bb^\tp\Sigmab\bb}.$$
Therefore, this robust model usually generates a more diversified portfolio than \eqref{eqn:TC}, as it also considers risk. Moreover, it takes into account the inaccuracy of return prediction. 

Similarly, the quadratic variable terms in the non-convex constraint in \eqref{eqn:TC_robust} mean that it is difficult to solve \eqref{eqn:TC_robust} directly, even with reformulation of the inner maximization problem. Accordingly, as in the proof of  Theorem~\ref{thm:TC_convex}, we first reformulate \eqref{eqn:TC_robust} into a non-convex problem without quadratic terms, and then further reformulate it into an SOCP that can be solved using off-the-shelf optimization solvers. Unlike the reformulation in Theorem~\ref{thm:TC_convex}, the square-root term in the SOCP (i.e., $\ell_2$ norm) requires us to make some mild assumptions to ensure the equivalence of the models. We begin by deriving the following lemma, which gives the Lipschitz constant of the $\ell_2$ norm.

\begin{lem} \label{lem:Lip_l2_norm}
Let $\sigma=\sqrt{\sum_{i=1}^m \sigma_i^2}$, where $\{\sigma_i^2\}$ are the diagonal entries of $\Sigmab=\Ub^\tp\Ub$. Then, for any $\{\bb_1,\bb_2\}\subseteq \R^m$, we have
$$\Big|\norms{\Ub\bb_1}_2- \norms{\Ub\bb_2}_2\Big| \leq \sigma\norm{\bb_1-\bb_2}_2.$$
\end{lem}

\begin{thm} \label{thm:TC_robust_SOCP}
Assume that $\max_i\{\xt_{t,i}\}>\kappa\sigma + \lambda$, where $\sigma=\sqrt{\sum_{i=1}^m \sigma_i^2}$ and $\{\sigma_i^2\}$ are the diagonal entries of $\Sigmab$. Consider the following program.
\begin{equation} \label{eqn:TC_robust_SOCP}
\begin{aligned}
& \underset{\bb\geq\zero}{\textup{maximize}}
& & \bb^\tp \xbt-\lambda\norms{\bbh_{t-1}-\bb}_1-\kappa\norms{\Ub\bb}_2\\
& \textup{subject to}
& & \bb^\tp\one + \gamma\norms{\bbh_{t-1}-\bb}_1 \leq 1,
\end{aligned}
\end{equation}
and let $\bb_{(3)}^*$ be an optimal solution. Then, $(w_{(1)}^*,\bb_{(1)}^*):=(\one^\tp\bb_{(3)}^*, \bb_{(3)}^*/\one^\tp\bb_{(3)}^*)$ is an optimal solution to \eqref{eqn:TC_robust}. 
\end{thm}

\begin{rem}
This condition ensures that $\kappa$ is sufficiently small, as it is easy to see that if $\kappa$ is too large, the optimal solution to \eqref{eqn:TC_robust_SOCP} is the zero vector. In this case, the equivalence cannot be established. The assumption $\max_i\{\xt_{t,i}\}>\kappa\sigma + \lambda$ is mild. As we use the most recent $m+1$ data to construct the shape parameter $\Sigmab$, the value of $\sigma$ is relatively small compared with the value of $\xbt$ (which typically fluctuates around $1$). In numerical experiments, the value of $\sigma$ generally ranges from $0.1$ to $0.3$ and the constraint in \eqref{eqn:TC_robust_SOCP} invariably achieves equality. Occasionally, if the chosen parameters are fairly large (which represents a fairly risk-averse investor), the condition may not hold; in these cases, we set the portfolio as $\bbh_{t-1}$. This is reasonable, as a risk-averse investor may not want to adjust the portfolio frequently. 
\end{rem}

\begin{rem}
If we do not have enough data to construct the shape parameter $\Sigmab\succ\zero$, we use $\bb_t=\bbh_{t-1}$ as our portfolio strategy. That is, we do not adjust the portfolio at all since perturbing the portfolio weights would incur transaction costs. We also note that there is a flexibility in choosing the shape parameter $\Sigmab$ by employing different covariance matrix estimates.
\end{rem}

In the reformulated model \eqref{eqn:TC_robust_SOCP}, the extra $\ell_2$ norm term in the objective function corresponds to the portfolio volatility (i.e., $\sqrt{\bb^\tp \Sigmab \bb}$). That is, the use of the ellipsoidal set takes into account the portfolio risk. Hence, the parameter $\kappa$ provides flexibility for adjustment of the risk attitude of the strategy. By combining this with the adaptive scheme for the choice of parameters discussed in the following section, we obtain a new proposal that allows us to construct a strategy that captures a satisfactory return under various market conditions (e.g., trending market conditions). This is because the strategy enables us to switch from passive risk-averse trading to active opportunity-seeking trading.

\section{Adaptive choice of model parameters} \label{sec:adaptive_scheme}
Despite the fact that model parameters in OLPS strategies play a vital role in the overall performance of such strategies, these parameters are usually tuned with reference to benchmark data sets. However, it may be imprudent to unthinkingly apply models with parameters tuned in this way to new data sets, as there is no guarantee that the default parameters will be optimal for application to non-test data sets. Accordingly, in this section we describe how the two parameters in our model---the size of the uncertainty set $\kappa$ and the rebalancing penalty $\lambda$, which are used to control the extent of robustness and the trading frequency, respectively---can be chosen in an adaptive manner. The pseudo code of the adaptive algorithms is provided in \ref{appdx:adaptive_scheme}.

\subsection{The adaptive scheme} \label{subsec:adaptive_scheme_description}
We first focus on the rebalancing penalty $\lambda$, which controls the trading frequency and hence the trading volume. This is a crucial consideration in the presence of transaction costs, as an unsuitable  $\lambda$ value may result in unsatisfactory portfolio performances as a result of over-active rebalancing or earning opportunities being missed due to passive holding. Thus, determining a suitable $\lambda$ value requires a sophisticated trade-off between exploiting opportunities and controlling transaction costs incurred from rebalancing a portfolio. 

The adaptive scheme takes a reasonably wide range of values of $\lambda$ as input: $\{\lambda_1,\dots,\lambda_L\}$. As in the CORN-U or CORN-K strategies \citep{Li_et_al:2011b}, we consider $L$ individual experts using our OLPS strategy with $L$ different choices of $\lambda$. At each period, we update the portfolio of each expert and the corresponding cumulative wealth. Therefore, at period $t$, we have $L$ different values for cumulative wealth for up to $t-1$ experts. We then decide on our investment strategies based on these $L$ experts. Designing an adaptive $\lambda$ scheme can be challenging; for instance, if experts with larger values of $\lambda$ perform better, simply increasing the current $\lambda$ (as in \citealp{Guo_et_al:2021}) may not be effective, as employing a large $\lambda$ generates a portfolio similar to the current portfolio. Therefore, our adaptive scheme shifts the entire portfolio to a completely new portfolio, i.e., we directly adopt the portfolio weights of one of the experts.

We decide which expert to follow by exploring the similarity of such a sequential decision-making process to the selection of the best (SB) problem (see, for instance, \citealp{Branke_et_al:2007, Fan_et_al:2016, Kim_Nelson:2006}). Solving such a problem requires the selection of the system---out of many other systems---that exhibits the best performance based on (sequential) samples. This mirrors the setting we have. Unlike the existing literature on selection of the best system, which attempts to develop different sampling methods that guarantee that the best system is selected with a high probability, we must sequentially select the ``best'' system, i.e., the expert with the greatest cumulative wealth. 

However, instead of switching in each period to the expert portfolio that has the best cumulative wealth in that period, we switch only when we are sufficiently ``confident.'' Specifically, for a given window size $W$ and expert $l$ that we are currently following, we compute the sample mean $\bar{S}^l_{t-1}$ and sample standard deviation $\hat{\sigma}^l_{t-1}$ based on this expert's cumulative wealth $\{S^l_{t-W},\dots,S^l_{t-1}\}$. If the moving average of another expert, say $\lambda_{l'}$, exceeds the upper limit $\bar{S}^l_{t-1}+z\hat{\sigma}^l_{t-1}$ (we select $z=1.96$ as the upper bandwidth of the standard normal distribution), we adjust the portfolio to that of expert $l'$, i.e., $\bb_t^{l'}$, in period $t$. Here, $z$ controls the trade-off between exploration (switching to the portfolio of another expert) and exploitation (remaining with the portfolio of the current expert), which is related to an investor's risk appetite. Moreover, we borrow the idea of an ``indifference zone'': if the difference between the cumulative wealth of another expert's portfolio and that of the current portfolio is less than $\delta$, we do not consider the other expert's portfolio as a better portfolio. This is useful in a stable market (i.e., when $\hat{\sigma}_t^l$ is small), which may result in active switching. However, in such a market, it is not rational to substantially change the portfolio positions as there are not significant differences between the performances of experts' portfolios. The overall idea of this adaptive scheme is summarized in Algorithm~\ref{algo:adaptive_lambda}, which we denote the SB scheme.

Each expert corresponds to a particular choice of $\lambda$ and thus, may execute our OLPS strategy with a flexible choice of $\kappa$. For example, one choice is to adopt the same SB scheme as in $\lambda$, i.e., switch to the portfolio of the expert that exhibits the best performance. Next, we develop an alternative scheme for $\kappa$ that can also achieve good performance in our numerical experiments.

In the developed model \eqref{eqn:TC_robust}, the parameter $\kappa$ controls the size of the uncertainty set. Thus, given Proposition~\ref{prop:inner_max}, $\kappa$ controls the trade-off between return and risk. Similarly, the adaptive scheme starts with a certain wide range of values of $\kappa$, say $\{\kappa_1,\dots,\kappa_N\}$. We consider $N$ individual experts using our OLPS strategy with $N$ difference choices of $\kappa$. At each period, we evaluate the portfolios of these experts and record their cumulative wealth. As our goal is to maximize cumulative wealth, we rank the experts in ascending order of their cumulative wealth. As in the CORN-K strategy, we set $\kappa$ for the next period as the average of the top-$K$ $\kappa$ values. Therefore, the value of $\kappa$ changes over time. Algorithm~\ref{algo:adaptive_kappa} summarizes this adaptive $\kappa$ scheme, which we denote the top-$K$ scheme. 

\begin{rem}
Our numerical experiments (see \ref{appdx:sensitivity}) indicate that selecting a $\lambda$ ranging from $\gamma$ to $100\gamma$ and a $\kappa$ ranging from $0.1$ to $10$ covers the best-performing combination of parameters. 
\end{rem}

\subsection{Advantages from the adaptive scheme} \label{subsec:adaptive_advantage}
Here, we summarize the advantages of our adaptive scheme strategy. First, the robustness parameter captures different extents of robustness, and hence the portfolio risk. Second, the adaptive schemes enable us to monitor the expert portfolios with a certain wide range of parameters. That is, we know how various parameters behave historically. In the design of the adaptive schemes, we focus on the experts' cumulative wealth and base our portfolio choice on this information. Therefore, we expect that we can capture a sudden market uptrend, as compared with experts with a high appetite for risk, experts with a low appetite for risk (i.e., a small $\kappa$ and $\lambda$) can more easily recognize such an uptrend as an opportunity to make gains. By adopting a greater appetite for risk, we may experience a larger portfolio volatility. In contrast, if the market experiences a sudden downtrend, experts with a higher risk appetite (i.e., a larger $\kappa$ and $\lambda$) can limit their trading volume and focus more on their portfolio risk. This may result in a better protection against such market downturns than active rebalancing. Therefore, by applying a combination of robustness and the adaptive scheme, our strategy has a good probability of yielding substantial cumulative wealth in practice. Moreover, unlike other methods, we need not specify a single value for the model. This means that our strategy can adapt to market conditions and maintain its robustness to various data sets, such that we can obtain satisfactory portfolio returns in most scenarios.

\section{Numerical experiments} \label{sec:numerics}

In this section, we perform extensive numerical experiments related to our model and schemes. In Section~\ref{subsec:experiment_description}, we describe the data sets we use and the experimental details. In Section~\ref{subsec:benchmark_data}, we apply our strategy and various other OLPS strategies to benchmark data sets, and compare these strategies’ performance using several metrics. Furthermore, we demonstrate the performance of our adaptive scheme with various values of $\kappa$ and $\lambda$. In Section~\ref{subsec:new_data}, we conduct similar experiments on some new data sets we have recently constructed.  We also reveal the importance of using adaptive schemes instead of the default parameters used in the literature.

\subsection{Experimental description} \label{subsec:experiment_description}
Following the vast literature on the OLPS problem (see, for instance, \citealp{Guo_et_al:2021,Huang_et_al:2016,Lai_et_al:2018,Li_et_al:2011,Li_et_al:2015,Li_et_al:2018}), we first perform our experiments on six benchmark data sets: DJIA \citep{Borodin_et_al:2004}, MSCI \citep{Li_Hoi:2018}, TSE \citep{Borodin_et_al:2004}, SP500 \citep{Borodin_et_al:2004}, NYSE-N \citep{Li_Hoi:2018} and NYSE-O \citep{Cover:1991}. These data sets are available online (see \citealp{Li_et_al:2016}). Table~\ref{table:data_set_benchmark} summarizes the six benchmark data sets and we refer readers to \cite{Li_Hoi:2018} for a detailed description.
\begin{table}[t]\centering\small
\ra{0.6}  
\caption{Summary of the six benchmark data sets} \label{table:data_set_benchmark}
\begin{tabular}{@{}ccccc@{}} \toprule
Data set & Region & Time period		           & Days & Number of assets\\
\midrule 
DJIA	    & USA	 & Jan 14, 2001 -- Jan 14, 2003    & 507 & 36 \\
MSCI	    & Global & Apr 1, 2006 -- Mar 31, 2010    & 1043 & 24 \\
TSE	        & CA	 & Jan 4, 1994 -- Dec 31, 1998    & 1259 & 88 \\
SP500       & USA	 & Jul 2, 1998 -- Jan 31, 2003    & 1276 & 25 \\
NYSE-N      & USA	 & Jan 1, 1985 -- Jun 30, 2010    & 6431 & 23 \\
NYSE-O      & USA	 & Jul 3, 1962 -- Dec 31, 1984    & 5651 & 36 \\
\bottomrule
\end{tabular}
\end{table}

We also test these strategies by applying them to three recent data sets that are summarized in Table~\ref{table:data_set_new}. The first data set is the SP500-21 data set, which comprises $50$ stocks we collect from the SP500 component stocks with the highest market capitalization in the US market. This data set covers the period from January 4, 2016, to August 30, 2021, during which time most of the stocks increased slightly in value. The second data set is the HSI data set, which comprises $30$ stocks we collect from the HSI component stocks with the highest market capitalization in the Hong Kong market. This data set covers the period from January 3, 2012, to August 30, 2021, during which time the HIS fluctuated and did not experience significant increases in value, i.e., most stocks increased slightly in value, aside from one particular stock that its values increased by more than 100 times.  The third data set is the CSI data set, which comprises $30$ stocks we collect from the CSI300 component stocks with the highest capitalization in the Chinese market. This data set covers the period from January 4, 2012, to August 30, 2021, during which time most of the stocks increased in value, to varying extents. We provide details of the new data sets in \ref{appdx:data_set}.
\begin{table}[t]\centering\small
\ra{0.6}  
\caption{Summary of the three new data sets} \label{table:data_set_new}
\begin{tabular}{@{}ccccc@{}} \toprule
Data set & Region & Time period		           & Days & Number of assets\\
\midrule 
SP500-21 & USA	   & Jan 4, 2016 -- Aug 30, 2021    & 1424 & 50 \\
HSI	    & HK     & Jan 3, 2012 -- Aug 30, 2021    & 2375 & 30 \\
CSI	    & CN	   & Jan 4, 2012 -- Aug 30, 2021    & 2345 & 30 \\
\bottomrule
\end{tabular}

\end{table}

We compare the values of various performance metrics of our strategy with those of several other OLPS strategies (see \ref{appdx:OLPS_para} for a summary of these strategies and their parameters). In Table~\ref{table:color}, we summarize the strategies into several classes. The performance metrics are related to the return and risks of a strategy, and are the cumulative wealth, the Sharpe ratio, the maximum drawdown and the Calmar ratio (see \ref{appdx:performance_metric} for a brief description of these metrics). We also provide numerical results for other metrics in \ref{appdx:compare_algorithms}.
\begin{table}[t]\centering\small
\ra{0.6}  
\caption{A brief classification of the strategies} \label{table:color}
\begin{tabular}{@{}lll@{}} \toprule
Color   & Description  & OLPS strategies\\
\midrule 
Red     & Our strategies            & RELP Adap (Top 5), RELP Adap (SB) \\
Blue    & Reference strategies      & UBAH, UCRP \\
Green   & Potential competitors     & RMR, TCO1, TCO2 \\
Purple  & Mean reversion strategies & AOLMA, CWMR-Var, OLMAR, PAMR  \\
Grey    & Other strategies          & Anticor, CORN-K, CORN-U, EG, ONS \\
\bottomrule
\end{tabular}
\end{table}

All numerical experiments are conducted in MATLAB on a computer with an AMD Ryzen 7 5800X 3.72 GHz CPU and 16 GB of memory. To solve our model as an SOCP, we employ the MATLAB solver \texttt{coneprog}, which implements the interior-point method developed by \cite{Andersen_et_al:2003}. The primal and dual feasibility tolerances are set to $10^{-5}$. The duality gap tolerance is also set to $10^{-5}$. We round the entries in the obtained solution with values less than $10^{-5}$ to zero, and then normalize the solution. This serves as the portfolio in the next period.

\subsection{Benchmark data sets} \label{subsec:benchmark_data}

\subsubsection{Comparison of portfolio strategies} \label{subsubsec:compare_algorithms}
In this section, we first compare our strategy with some existing OLPS strategies, as detailed in \ref{appdx:OLPS_para}. In our model, we first set $\lambda = 10\gamma$, similar to \cite{Li_et_al:2018}, and we set $\kappa$ using the two adaptive schemes we discuss in Section~\ref{subsec:adaptive_scheme_description}, namely (a) the top-$K$ updating scheme with $K=5$, and (b) the SB scheme with $\delta = 0$ and $W=5$. We denote the former scheme ``RELP Adap (Top 5) '' and the latter scheme ``RELP Adap (SB).'' We use $31$ values for $\kappa$ that are in a log-scale between $0.1$ and $10$, inclusive. We use $\gamma\in\{0,0.002,0.005\}$, which corresponds to $0\%$,  $0.2\%$ and $0.5\%$ transaction costs, respectively.



Different strategies are often compared based on the ranking of their performance metrics. However, the actual values of the performance metrics are more important, as they can better reflect the performance of a strategy than rankings. Therefore, we make use of a relative ranking, which is defined as
$$\text{relative ranking}=\frac{\text{stat}-\text{stat}_\text{worst}}{\text{stat}_\text{best}-\text{stat}_\text{worst}},$$
where $\text{stat}_\text{best}$ and $\text{stat}_\text{worst}$ are the best-performance metric value and the worst-performance metric value over the testing strategies, respectively. Therefore, the relative ranking  is a value between $0$ and $1$: it takes a value of $1$ if a strategy yields the best performance metric and takes a value of $0$ if the strategy yields the worst performance metric. This also takes into account how far one metric deviates from the best (or the worst) metric among different strategies. Next, we present box plots that summarize the relative rankings (for various performance metrics) of the strategies with different transaction costs and applied to various data sets. In this experiment, we use three transaction costs and six benchmark data sets, which generates $18$ observations for each strategy (see Table~\ref{table:color} for the strategies).

\begin{figure}[t]
\centering
\begin{subfigure}{0.475\textwidth}
\centering
\includegraphics[scale=0.57]{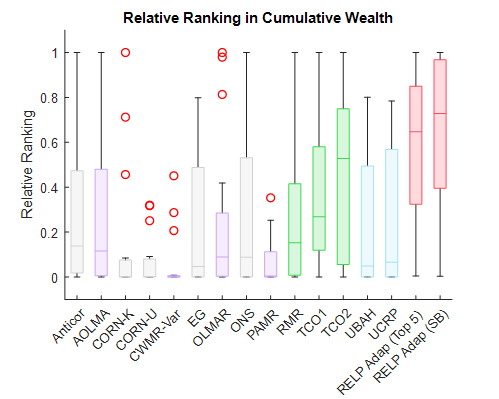}
\caption{Cumulative wealth} \label{fig:benchmark_wealth_summary}
\end{subfigure}
\hfill
\begin{subfigure}{0.475\textwidth}
\centering
\includegraphics[scale=0.57]{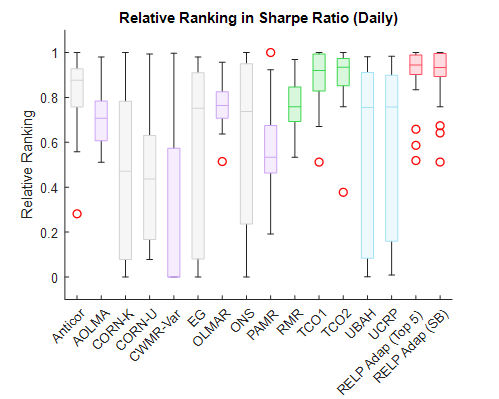}
\caption{Sharpe ratio (daily)} \label{fig:benchmark_sharpe_summary}
\end{subfigure}
\caption{Relative rankings on cumulative wealth and Sharpe ratio} \label{fig:benchmark_wealth_sharpe_summary}
\end{figure}

First, we discuss the cumulative wealth performance of the strategies. Figure~\ref{fig:benchmark_wealth_summary} shows the relative rankings of the strategies' cumulative wealth. Both of the RELP Adap strategies outperform many existing OLPS strategies (e.g., with respect to the median of the rankings), and are the top-ranking strategies. TCO1 and TCO2 are two other strategies that account for transaction costs \cite{Li_et_al:2018}, and thus also yield satisfactory levels of cumulative wealth. The RMR strategy, which employs a robust estimator, may also occasionally yield high cumulative wealth. These results demonstrate that the RELP Adap strategies capture the best returns in most scenarios. We attribute this to our adaptive scheme, as it covers a wide range of risk attitudes associated with $\kappa$.

%

\begin{figure}[t] 
\hspace{-15mm}
\includegraphics[scale=0.58]{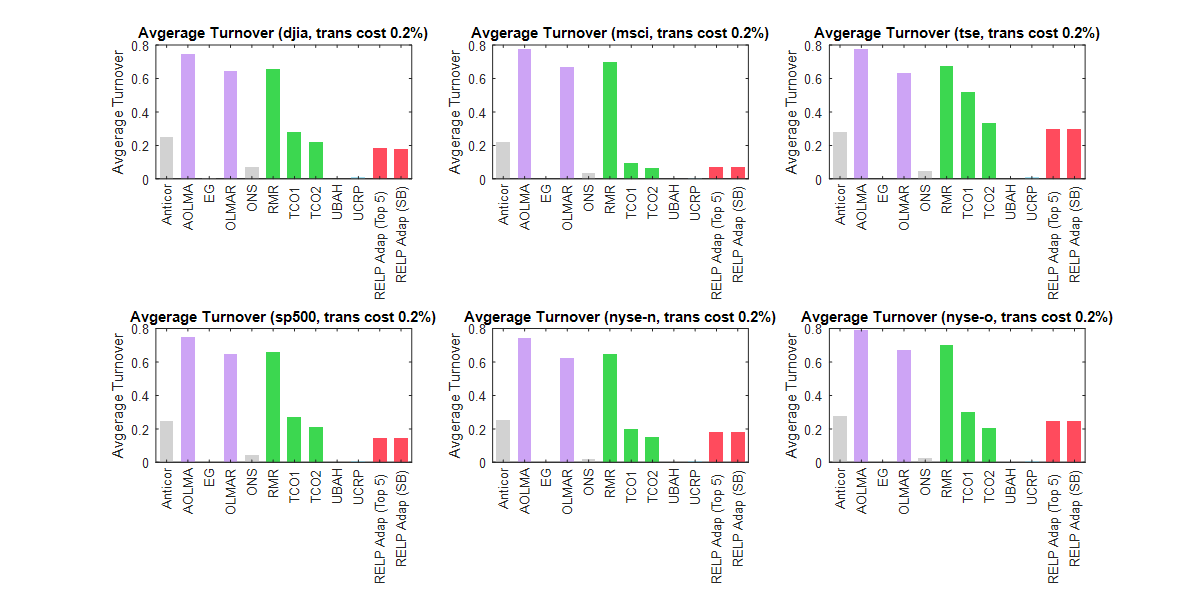}
\caption{Average turnover of various strategies applied to the benchmark data sets with $\gamma=0.2\%$ } \label{fig:benchmark_turnover_gamma_2}
\end{figure}

To illustrate the trade-off between active rebalancing and transaction costs, we also compute the average turnover of the strategies, which is defined as
\begin{equation} \label{eqn:average_turnover}
 AT=\frac{1}{2n}\sum_{t=1}^n \norm{\bbh_{t-1}-w_{t-1}\bb_t}_1,
\end{equation}
which captures the actual change in portfolio weight \citep{Li_et_al:2018}. We present the results for selected strategies in Figure~\ref{fig:benchmark_turnover_gamma_2}. The strategies that do not penalize the change in portfolio (e.g., the OLMAR, AOLMA, and RMR strategies) usually generate huge turnovers. In contrast, the reference strategies (UBAH and UCRP) generate tiny turnovers. In addition, the EG and ONS strategies, which include a regularization term on the change in portfolio and on the portfolio itself, yield a tiny turnover. Moreover, when using the default parameters, which is a more conservative approach, the overall performance of the EG and ONS strategies is similar to that of the reference strategies. The RELP Adap and TCO strategies both maintain a small turnover and generate outstanding cumulative wealth, indicating that these strategies strike a good balance between exploiting earning opportunities and minimizing transaction costs. 

Figure~\ref{fig:benchmark_sharpe_summary} shows the relative rankings of strategies in terms of daily Sharpe ratios, which represent the (portfolio volatility) risk-adjusted return. Again, our RELP Adap strategies outperform many other strategies, aside from the TCO strategies. This may be due to the fact that an increase in the overall terminal wealth, as observed in Figure~\ref{fig:benchmark_wealth_summary}, is accompanied by an increase in portfolio volatility. The RMR strategy has outstanding performance in general. These results show that our RELP Adap strategies achieve satisfactory Sharpe ratios.

\begin{figure}[t] 
\centering
\begin{subfigure}{0.475\textwidth}
\centering
\includegraphics[scale=0.57]{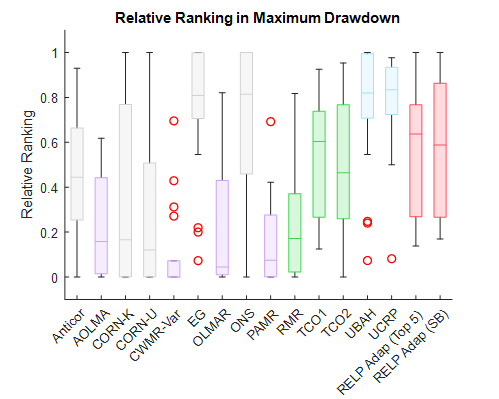}
\caption{Maximum drawdown} \label{fig:benchmark_mdd_summary}
\end{subfigure}
\hfill
\begin{subfigure}{0.475\textwidth}
\centering
\includegraphics[scale=0.57]{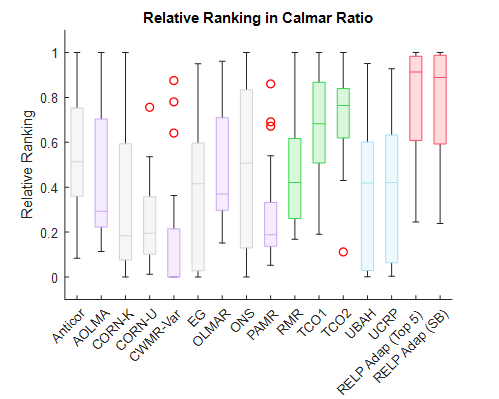}
\caption{Calmar ratio} \label{fig:benchmark_calmar_summary}
\end{subfigure}
\caption{Relative rankings of strategies in terms of maximum drawdown and Calmar ratio} \label{fig:benchmark_mdd_calmar_summary}
\end{figure}

Next, we consider maximum drawdown, which captures the worst-case decrease in returns during the investment horizon. Such a scenario is undesirable, as it reduces the investment capital by a significant amount. Figure~\ref{fig:benchmark_mdd_summary} summarizes the results. The active strategies (i.e., EG, ONS, TCO1 and TCO2), which take transaction costs into account, and the reference strategies (i.e., UBAH and UCRP), usually produce a smaller maximum drawdown than the other strategies. Our RELP Adap strategies generally produce a similar or smaller maximum drawdown than the TCO and RMR strategies. This may be attributable to the design of our adaptive scheme, which enables it to follow the expert who has the best cumulative wealth. If a market downtrend occurs, our strategies may become more risk-averse than in a normal market, and thus avoid active trading. This demonstrates the capability of our strategies to protect against unfavorable scenarios. Figure~\ref{fig:benchmark_calmar_summary} shows the results in terms of the Calmar ratio, which is the maximum drawdown-adjusted return. The cumulative wealth and maximum drawdown achieved by our RELP Adap strategies demonstrate that they substantially outperform the other OLPS strategies. 

\subsubsection{Comparison of performance  of adaptive schemes} \label{subsubsec:adaptive_schemes}
In this section, we compare the performance of our adaptive schemes with that of several other adaptive schemes. First, we compare the performance of adaptive schemes with $31$ values for $\kappa$ from a log-scale between $0.1$ and $10$ (inclusive) as the input. We consider four different settings: $\gamma\in\{0.2\%,\,0.5\%\}$ and $\lambda\in\{10\gamma,\,50\gamma\}$. Similar to Section~\ref{subsubsec:compare_algorithms}, we use relative rankings and construct box plots based on $24$ observations.


Table~\ref{table:adaptive_kappa} summarizes the compared adaptive schemes. Schemes (a) to (d) are the updating methods that are used to compute a new $\kappa$ for the next period from an expert's information. The top-$K$ scheme developed in Section~\ref{sec:adaptive_scheme} corresponds to schemes (a) to (c) with $K\in\{1,3,5\}$. Scheme (d) captures both the most recent information $\kappa_\text{best}$ and the past information $\kappa_\text{prev}$. Hence, it is a weighted average of all previous $\kappa_\text{best}$ with exponentially decaying weights. In contrast, schemes (e) to (j) are methods that make use of the expert's portfolios directly. The adaptive scheme for $\lambda$ (which is also used for $\kappa$)  corresponds to schemes (e) and (f). In schemes (g) to (j), the portfolio is updated as a weighted average of an expert's portfolios. In particular, scheme (i) is adopted in the CORN-K strategy.
\begin{table}[t]\centering\small
\ra{0.6}  
\caption{Various compared adaptive $\kappa$ schemes.} \label{table:adaptive_kappa}
\begin{tabular}{@{}cl@{}} \toprule
Index & Description\\
\midrule 
(a)     & Updating scheme based on the (geometric) average of five $\kappa$'s with best cumulative return \\
(b)	 & Updating scheme based on the (geometric) average of three $\kappa$'s with best cumulative return \\
(c)	 & Updating scheme based on the $\kappa$ with best cumulative return, i.e., $\kappa_\text{best}$ \\
(d)     & Updating scheme based on $0.5\kappa_\text{best}+0.5\kappa_\text{prev}$ where $\kappa_\text{prev}$ is the $\kappa$ used in the previous period \\
(e)     & SB scheme with $\delta = 0$ \\
(f)     & SB scheme with $\delta = 0.2$ \\
(g)     & Use the portfolio with the best cumulative return \\
(h)	 & Use the average of five portfolios with best cumulative return \\
(i)     & Use the cumulative-return weighted average of five portfolios with best cumulative return \\
(j)     & Use the average of all the $\kappa$ portfolios \\
\bottomrule
\end{tabular}
\end{table}

\begin{figure}[t]
\centering
\begin{subfigure}{0.475\textwidth}
\centering
\includegraphics[scale=0.57]{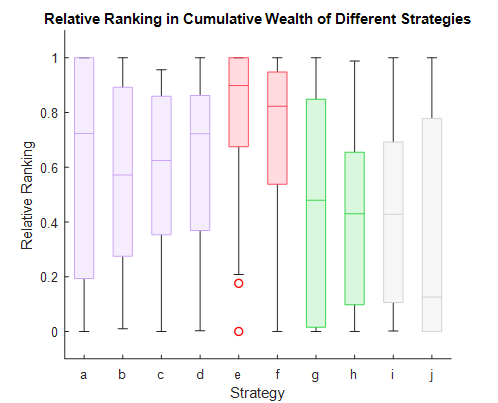}
\caption{Cumulative wealth} \label{fig:adap_kappa_wealth_summary}
\end{subfigure}
\hfill
\begin{subfigure}{0.475\textwidth}
\centering
\includegraphics[scale=0.57]{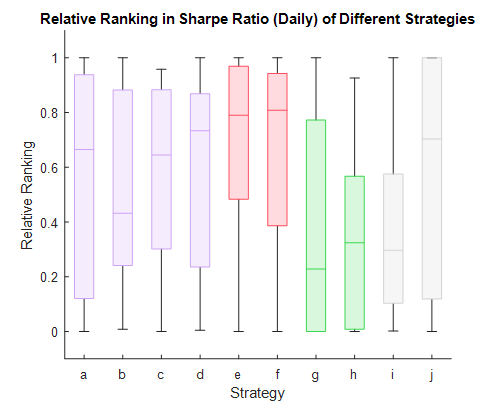}
\caption{Sharpe ratio (daily)} \label{fig:adap_kappa_sharpe_summary}
\end{subfigure}
\caption{Relative rankings of various adaptive $\kappa$ schemes in terms of cumulative wealth and Sharpe ratio} \label{fig:adap_kappa_wealth_sharpe_summary}
\end{figure}

Figure~\ref{fig:adap_kappa_wealth_summary} shows the relative rankings in terms of cumulative wealth for various adaptive schemes based on $\kappa$ values. Of the updating schemes, i.e., schemes (a) to (d), scheme (a) usually yields the best cumulative wealth. As $\kappa_\text{best}$ may be noisy---in the sense that its value frequently changes---scheme (a) tends to perform well, as it adopts a local averaging update. Of the remaining schemes, scheme (j) usually has the worst performance; this is logical, as scheme (j) is an averaging scheme, which dilutes the effect of active balancing. Thus, it cannot capture huge returns from a market uptrend. Schemes (h) and (i) have similar performances, as they both use the five best-performing portfolios, albeit with different assigned weights. In scheme (g), $\kappa_\text{best}$ may exhibit frequent jumps (as explained previously), which would induce huge transaction costs. In schemes (e) and (f), we switch the expert (i.e., change the monitoring expert) only when another expert sufficiently dominates the current expert. Therefore, schemes (e) and (f) usually have a small turnover, and their portfolio weights are dramatically changed only when we are sufficiently confident. Overall, schemes (a) and (e) generally produce a better cumulative wealth than the other schemes. Figure~\ref{fig:adap_kappa_sharpe_summary} shows the results in terms of Sharpe ratios, which indicates that the relative performances of schemes are similar to those for various $\kappa$ values, except for scheme (j). This is because scheme (j) averages over the expert's entire portfolio, which stabilizes the change in portfolio and leads to a small portfolio volatility.


\begin{figure}[t]  
\centering
\begin{subfigure}{0.475\textwidth}
\centering
\includegraphics[scale=0.57]{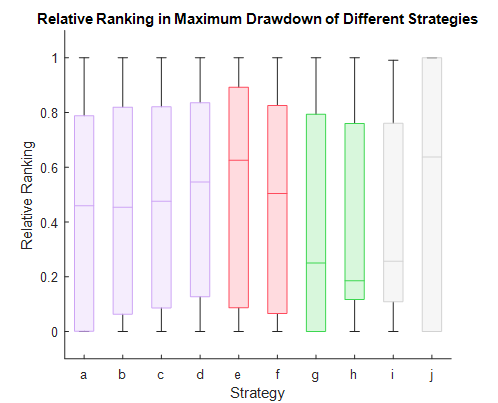}
\caption{Maximum drawdown} \label{fig:adap_kappa_mdd_summary}
\end{subfigure}
\hfill
\begin{subfigure}{0.475\textwidth}
\centering
\includegraphics[scale=0.57]{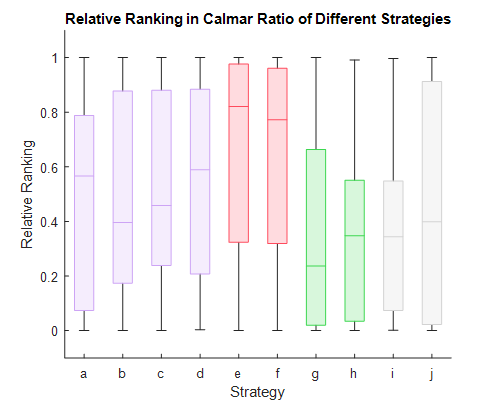}
\caption{Calmar ratio} \label{fig:adap_kappa_calmar_summary}
\end{subfigure}
\caption{Relative rankings of various adaptive $\kappa$ schemes in terms of maximum drawdown and Calmar ratios} \label{fig:adap_kappa_mdd_calmar_summary}
\end{figure}

Figures~\ref{fig:adap_kappa_mdd_summary} and \ref{fig:adap_kappa_calmar_summary} show the performances of schemes in terms of maximum drawdown and Calmar ratios, respectively. The performances of the updating schemes (a) to (d) are generally similar. In contrast, scheme (j) generates the smallest or the largest maximum drawdown or Calmar ratio across different settings. This may not be desirable, as it may indicate that a strategy based on scheme (j) has unstable performance. Schemes (g) to (i) behave similarly and appear to be under-performing, as they are outperformed by both schemes (e) and (f). In addition, in terms of Calmar ratios, schemes (e) and (f) have the best performance.

%

Next, we compare various adaptive schemes in terms of $\lambda$. We consider six different settings: $\gamma\in\{0.2\%,\, 0.5\%\}$ and $\kappa\in\{0.2,\,0.5,\,1\}$. Therefore, we have $36$ data points in total. We take $\lambda=\delta\gamma$, where $\delta$ takes $31$ values in a log-scale between $1$ and $100$ (inclusive) as its input. Table~\ref{table:adaptive_lambda} summarizes and compares all of the adaptive schemes, which are similar to schemes (e) to (j) in Table~\ref{table:adaptive_kappa}. 

%
\begin{table}[t]\centering\small
\ra{0.6}  
\caption{Various compared adaptive $\lambda$ schemes} \label{table:adaptive_lambda}
\begin{tabular}{@{}cl@{}} \toprule
Index & Description\\
\midrule 
(i)     & SB scheme with $\delta = 0$ \\
(ii)    & SB scheme with $\delta = 0.2$ \\
(iii)   & Use the portfolio with the best cumulative return \\
(iv)	   & Use the average of five portfolios with the best cumulative returns \\
(v)     & Use the cumulative-return weighted average of five portfolios with the best cumulative returns \\
(vi)    & Use the average of all the $\lambda$ portfolios \\
\bottomrule
\end{tabular}
\end{table}

\begin{figure}[t] 
\centering
\begin{subfigure}{0.475\textwidth}
\centering
\includegraphics[scale=0.57]{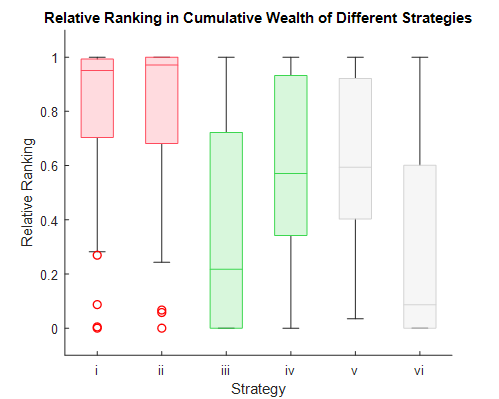}
\caption{Cumulative wealth} \label{fig:adap_lambda_wealth_summary}
\end{subfigure}
\hfill
\begin{subfigure}{0.475\textwidth}
\centering
\includegraphics[scale=0.57]{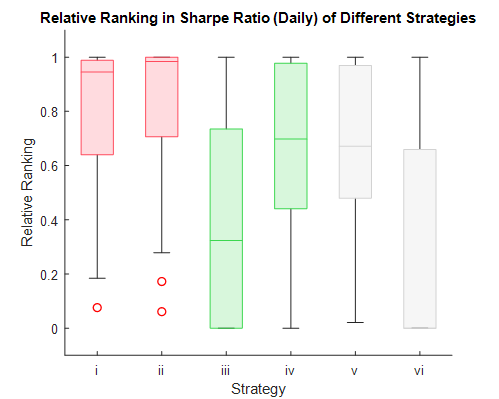}
\caption{Sharpe ratio (daily)} \label{fig:adap_lambda_sharpe_summary}
\end{subfigure}
\caption{Relative rankings of various adaptive $\lambda$ schemes in terms of cumulative wealth and Sharpe ratios} \label{fig:adap_lambda_wealth_sharpe_summary}
\end{figure}

Figure~\ref{fig:adap_lambda_wealth_summary} shows the relative rankings of schemes in terms of cumulative wealth. Schemes (i) and (ii) generate much better cumulative wealth than the other schemes. This demonstrates the superiority of the our adaptive scheme. In scheme (iii), the best $\lambda$ changes rapidly, which leads to huge transaction costs and hence poor performance. While schemes (iv) and (v) are based on the top five experts' portfolios, they do not outperform schemes (a) and (b) in general. Finally, scheme (vi), which takes an average over all of the experts' portfolios, yields the poorest cumulative wealth. This is reasonable, as this approach of scheme (vi) means that it considers some poorly performing portfolios. We observe similar patterns in the Sharpe ratios in Figure~\ref{fig:adap_lambda_sharpe_summary}.

\begin{figure} 
\centering
\begin{subfigure}{0.475\textwidth}
\centering
\includegraphics[scale=0.57]{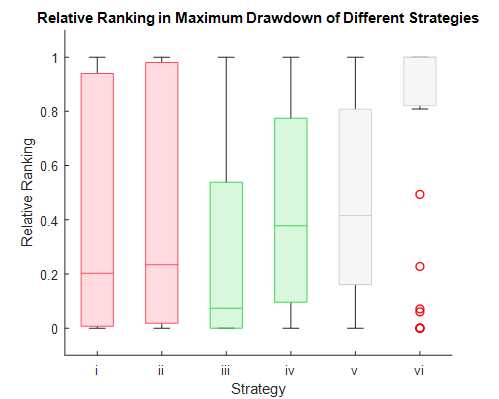}
\caption{Maximum drawdown} \label{fig:adap_lambda_mdd_summary}
\end{subfigure}
\hfill
\begin{subfigure}{0.475\textwidth}
\centering
\includegraphics[scale=0.57]{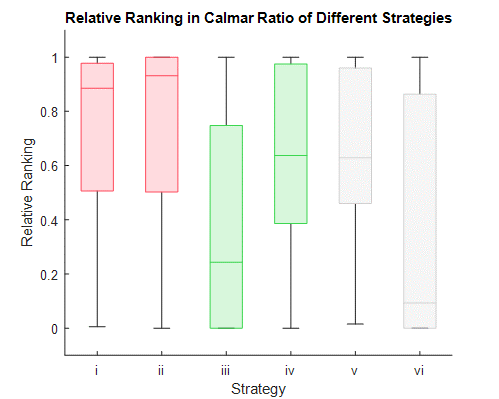}
\caption{Calmar ratio} \label{fig:adap_lambda_calmar_summary}
\end{subfigure}
\caption{Relative rankings of various adaptive $\lambda$ schemes in  terms of maximum drawdown and Calmar ratios} \label{fig:adap_lambda_mdd_calmar_summary}
\end{figure}

Figure~\ref{fig:adap_lambda_mdd_summary} compares schemes in terms of maximum drawdown. The maximum drawdown from schemes (i) and (ii) deviates greatly from that of the other schemes. This is reasonable, as these schemes monitor experts with small $\lambda$, who try to exploit the market uptrend. However, from Figure~\ref{fig:adap_lambda_mdd_calmar_summary}, we observe that one scheme generally outperforms the others in terms of the Calmar ratio. That is, similar to previous observations, scheme (iii) gives the poorest performance whereas schemes (iv) and (v) behave similarly, and better than scheme (iii). Scheme (vi) takes an average over all experts' portfolios, and yields the best maximum drawdown in many settings. However, this is not commensurate with its poor cumulative return, as shown in Figure~\ref{fig:adap_lambda_mdd_calmar_summary}.

Finally, we apply our model using the following fully adaptive schemes: (1) the SB scheme to both $\kappa$ and $\lambda$ with $\delta=0$ (the RELP-Adap-1 strategy); and (2) the SB scheme to $\lambda$ with $\delta=0$ and the top-5 scheme to $\kappa$ (the RELP-Adap-2 strategy). Tables~\ref{table:benchmark_cumulative_wealth} to \ref{table:benchmark_calmar} provide the cumulative wealth, Sharpe ratios and Calmar ratios for various data sets with $\gamma=0.2\%$. For brevity, we select several existing strategies for comparison. The color intensity gives the relative performances of the OLPS strategies (where a higher intensity indicates a better performance). Overall, the RELP strategies perform well when applied to various data sets, as most of the values of these strategies' statistics are the two best overall. This demonstrates that our RELP strategies achieve superior performances. We note that our strategies do not perform well when applied to the DJIA data set, which is also true for many OLPS strategies \citep{Li_et_al:2012}. This may be due to a lack of general trend in the assets in this data set, leading to differences between experts' performances. This results in a rapid change in the choice of expert, resulting in large transaction costs. In this case, we can maintain a return close to the market strategy. One way to rectify such a situation is to select a sparser scale of parameters (rather than a scale containing $31$ different values of $\lambda$ and $\kappa$ over a relatively small range), so that the choice of expert does not change rapidly. This provides flexibility to our RELP strategies, as choice is related to an investor's preferences. Nevertheless, our RELP strategies outperform many other strategies. In \ref{appdx:full_adaptive_benchmark}, we also present similar results that are obtained when $\gamma=0.5\%$. In the TSE data set, which involves the largest number of assets, the average solution time of the SOCP for various combinations of parameters is $0.042$ s. Moreover, when applying the adaptive schemes (with multiple parameter combinations), parallel computing can be utilized to accelerate the computation. Hence, our RELP strategies are time efficient from a practical perspective.

\begin{table}[t]\small
\centering
\ra{0.6}
\caption{Comparison of cumulative wealth generated by schemes applied to six benchmark data sets} \label{table:benchmark_cumulative_wealth}
\begin{tabular}{@{}rrrrrrr@{}} \toprule
Strategy   ($\gamma=0.2\%$) & DJIA                           & MSCI                           & TSE                             & SP500                          & NYSE-N                           & NYSE-O                           \\ \cmidrule{2-7}
EG                          & \cellcolor[HTML]{F1F7FC}0.7601 & \cellcolor[HTML]{FAFCFE}0.9031 & \cellcolor[HTML]{FEFFFF}1.6009  & \cellcolor[HTML]{F1F7FC}1.3218 & \cellcolor[HTML]{FFFFFF}1.76E+01 & \cellcolor[HTML]{FFFFFF}1.40E+01 \\
OLMAR                       & \cellcolor[HTML]{FFFFFF}0.5978 & \cellcolor[HTML]{FAFCFE}0.9015 & \cellcolor[HTML]{FBFDFE}2.6383  & \cellcolor[HTML]{FBFDFE}0.6377 & \cellcolor[HTML]{FFFFFF}4.40E+01 & \cellcolor[HTML]{F4F8FC}2.02E+10 \\
ONS                         & \cellcolor[HTML]{BDD7EE}1.3322 & \cellcolor[HTML]{FFFFFF}0.7416 & \cellcolor[HTML]{FFFFFF}1.2782  & \cellcolor[HTML]{DFECF7}2.6745 & \cellcolor[HTML]{FFFFFF}1.28E+01 & \cellcolor[HTML]{FFFFFF}5.98E+01 \\
RMR                         & \cellcolor[HTML]{F8FBFE}0.6773 & \cellcolor[HTML]{FBFDFE}0.8805 & \cellcolor[HTML]{E6F0F9}9.0662  & \cellcolor[HTML]{FFFFFF}0.2723 & \cellcolor[HTML]{FFFFFF}1.57E+01 & \cellcolor[HTML]{F0F6FB}2.72E+10 \\
TCO1                        & \cellcolor[HTML]{ECF4FA}0.8159 & \cellcolor[HTML]{D9E8F6}1.8814 & \cellcolor[HTML]{DDEBF7}11.7100 & \cellcolor[HTML]{FAFCFE}0.6676 & \cellcolor[HTML]{FBFCFE}1.49E+04 & \cellcolor[HTML]{ECF4FB}3.30E+10 \\
TCO2                        & \cellcolor[HTML]{CCE0F2}1.1718 & \cellcolor[HTML]{DAE9F6}1.8657 & \cellcolor[HTML]{BDD7EE}21.2827 & \cellcolor[HTML]{DFECF7}2.7053 & \cellcolor[HTML]{F9FCFE}1.80E+04 & \cellcolor[HTML]{FFFFFF}6.42E+08 \\
UBAH                        & \cellcolor[HTML]{F1F7FC}0.7628 & \cellcolor[HTML]{FAFCFE}0.9045 & \cellcolor[HTML]{FEFFFF}1.6097  & \cellcolor[HTML]{F1F7FC}1.3390 & \cellcolor[HTML]{FFFFFF}1.80E+01 & \cellcolor[HTML]{FFFFFF}1.45E+01 \\
UCRP                        & \cellcolor[HTML]{EDF5FB}0.7996 & \cellcolor[HTML]{FAFCFE}0.9092 & \cellcolor[HTML]{FFFFFF}1.5363  & \cellcolor[HTML]{EEF5FB}1.5818 & \cellcolor[HTML]{FFFFFF}2.70E+01 & \cellcolor[HTML]{FFFFFF}2.37E+01 \\
RELP-Adap-1                 & \cellcolor[HTML]{FDFEFF}0.6268 & \cellcolor[HTML]{BDD7EE}2.6984 & \cellcolor[HTML]{DDEAF7}11.7940 & \cellcolor[HTML]{BDD7EE}5.2002 & \cellcolor[HTML]{DBE9F6}1.10E+05 & \cellcolor[HTML]{BDD7EE}1.14E+11 \\
RELP-Adap-2                 & \cellcolor[HTML]{F3F8FC}0.7372 & \cellcolor[HTML]{CFE2F3}2.1769 & \cellcolor[HTML]{E9F2FA}8.2372  & \cellcolor[HTML]{BED8EF}5.1258 & \cellcolor[HTML]{BDD7EE}1.97E+05 & \cellcolor[HTML]{C3DBF0}1.05E+11\\
\bottomrule
\end{tabular}
\end{table}
\begin{table}[t]\small
\centering
\ra{0.6}
\caption{Comparison of Sharpe ratios of schemes applied to six benchmark data sets} 
\label{table:benchmark_sharpe}
\begin{tabular}{@{}rrrrrrr@{}} \toprule
Strategy   ($\gamma=0.2\%$) & DJIA                            & MSCI                            & TSE                            & SP500                           & NYSE-N                         & NYSE-O                         \\ \cmidrule{2-7}
EG                          & \cellcolor[HTML]{FFFFFF}-0.0378 & \cellcolor[HTML]{FBFDFE}-0.0086 & \cellcolor[HTML]{E9F2FA}0.0307 & \cellcolor[HTML]{E4EFF8}0.0118  & \cellcolor[HTML]{F2F7FC}0.0315 & \cellcolor[HTML]{FFFFFF}0.0376 \\
OLMAR                       & \cellcolor[HTML]{EEF5FB}-0.0199 & \cellcolor[HTML]{F0F6FC}0.0021  & \cellcolor[HTML]{E0EDF7}0.0382 & \cellcolor[HTML]{EDF4FB}0.0020  & \cellcolor[HTML]{F5F9FD}0.0297 & \cellcolor[HTML]{CEE2F3}0.1308 \\
ONS                         & \cellcolor[HTML]{BDD7EE}0.0300  & \cellcolor[HTML]{FFFFFF}-0.0133 & \cellcolor[HTML]{FFFFFF}0.0121 & \cellcolor[HTML]{C2DAF0}0.0484  & \cellcolor[HTML]{FFFFFF}0.0223 & \cellcolor[HTML]{F6FAFD}0.0559 \\
RMR                         & \cellcolor[HTML]{E6F0F9}-0.0119 & \cellcolor[HTML]{F1F7FC}0.0010  & \cellcolor[HTML]{CDE1F2}0.0545 & \cellcolor[HTML]{FFFFFF}-0.0176 & \cellcolor[HTML]{FBFDFE}0.0251 & \cellcolor[HTML]{CEE1F3}0.1316 \\
TCO1                        & \cellcolor[HTML]{E6F0F9}-0.0114 & \cellcolor[HTML]{D2E4F4}0.0325  & \cellcolor[HTML]{C8DEF1}0.0588 & \cellcolor[HTML]{F4F9FD}-0.0053 & \cellcolor[HTML]{C7DDF1}0.0599 & \cellcolor[HTML]{BDD7EE}0.1619 \\
TCO2                        & \cellcolor[HTML]{C8DEF1}0.0189  & \cellcolor[HTML]{D3E5F4}0.0311  & \cellcolor[HTML]{BDD7EE}0.0672 & \cellcolor[HTML]{CEE1F3}0.0358  & \cellcolor[HTML]{CADFF2}0.0580 & \cellcolor[HTML]{CEE2F3}0.1304 \\
UBAH                        & \cellcolor[HTML]{FFFFFF}-0.0373 & \cellcolor[HTML]{FBFDFE}-0.0085 & \cellcolor[HTML]{E8F2FA}0.0313 & \cellcolor[HTML]{E4EFF8}0.0124  & \cellcolor[HTML]{F1F7FC}0.0318 & \cellcolor[HTML]{FFFFFF}0.0386 \\
UCRP                        & \cellcolor[HTML]{F7FAFD}-0.0292 & \cellcolor[HTML]{FAFCFE}-0.0076 & \cellcolor[HTML]{EEF5FB}0.0268 & \cellcolor[HTML]{DBE9F6}0.0216  & \cellcolor[HTML]{ECF3FA}0.0357 & \cellcolor[HTML]{F8FBFE}0.0520 \\
RELP-Adap-1                 & \cellcolor[HTML]{F0F6FC}-0.0224 & \cellcolor[HTML]{BDD7EE}0.0531  & \cellcolor[HTML]{C7DDF1}0.0591 & \cellcolor[HTML]{BED8EF}0.0533  & \cellcolor[HTML]{C2DAF0}0.0636 & \cellcolor[HTML]{C8DEF1}0.1418 \\
RELP-Adap-2                 & \cellcolor[HTML]{E4EFF8}-0.0098 & \cellcolor[HTML]{C8DEF1}0.0426  & \cellcolor[HTML]{CEE2F3}0.0533 & \cellcolor[HTML]{BDD7EE}0.0536  & \cellcolor[HTML]{BDD7EE}0.0665 & \cellcolor[HTML]{C8DEF1}0.1428\\
\bottomrule
\end{tabular}
\end{table}
\begin{table}[t]\small
\centering
\ra{0.6}
\caption{Comparison of Calmar ratios of schemes applied to six benchmark data sets} 
\label{table:benchmark_calmar}
\begin{tabular}{@{}rrrrrrr@{}} \toprule
Strategy   ($\gamma=0.2\%$) & DJIA                            & MSCI                            & TSE                            & SP500                           & NYSE-N                         & NYSE-O                         \\ \cmidrule{2-7}
EG                          & \cellcolor[HTML]{FFFFFF}-0.3301 & \cellcolor[HTML]{FBFDFE}-0.0376 & \cellcolor[HTML]{F1F7FC}0.3253 & \cellcolor[HTML]{EBF3FA}0.1213  & \cellcolor[HTML]{F2F7FC}0.2234 & \cellcolor[HTML]{FFFFFF}0.2942 \\
OLMAR                       & \cellcolor[HTML]{FEFFFF}-0.3182 & \cellcolor[HTML]{FBFCFE}-0.0364 & \cellcolor[HTML]{F7FBFD}0.2306 & \cellcolor[HTML]{F8FBFE}-0.1443 & \cellcolor[HTML]{FAFCFE}0.1613 & \cellcolor[HTML]{C8DEF1}3.5645 \\
ONS                         & \cellcolor[HTML]{BDD7EE}0.4448  & \cellcolor[HTML]{FFFFFF}-0.0997 & \cellcolor[HTML]{FFFFFF}0.1016 & \cellcolor[HTML]{C7DDF1}0.8494  & \cellcolor[HTML]{FFFFFF}0.1141 & \cellcolor[HTML]{F8FBFE}0.7128 \\
RMR                         & \cellcolor[HTML]{F8FBFE}-0.2476 & \cellcolor[HTML]{FBFDFE}-0.0440 & \cellcolor[HTML]{DDEBF7}0.6429 & \cellcolor[HTML]{FFFFFF}-0.2895 & \cellcolor[HTML]{FFFFFF}0.1148 & \cellcolor[HTML]{C8DEF1}3.5607 \\
TCO1                        & \cellcolor[HTML]{F3F8FC}-0.1845 & \cellcolor[HTML]{DEEBF7}0.3302  & \cellcolor[HTML]{CEE2F3}0.8722 & \cellcolor[HTML]{F8FBFD}-0.1327 & \cellcolor[HTML]{D1E3F3}0.4895 & \cellcolor[HTML]{BDD7EE}4.1603 \\
TCO2                        & \cellcolor[HTML]{D1E3F3}0.2193  & \cellcolor[HTML]{E1EDF8}0.2909  & \cellcolor[HTML]{C2DAF0}1.0706 & \cellcolor[HTML]{D4E5F4}0.5836  & \cellcolor[HTML]{D1E3F3}0.4880 & \cellcolor[HTML]{D0E2F3}3.1018 \\
UBAH                        & \cellcolor[HTML]{FFFFFF}-0.3266 & \cellcolor[HTML]{FBFDFE}-0.0370 & \cellcolor[HTML]{F1F7FC}0.3301 & \cellcolor[HTML]{EBF3FA}0.1294  & \cellcolor[HTML]{F2F7FC}0.2241 & \cellcolor[HTML]{FFFFFF}0.3039 \\
UCRP                        & \cellcolor[HTML]{FBFDFE}-0.2742 & \cellcolor[HTML]{FAFCFE}-0.0352 & \cellcolor[HTML]{F5F9FD}0.2649 & \cellcolor[HTML]{E2EEF8}0.2961  & \cellcolor[HTML]{F3F8FC}0.2124 & \cellcolor[HTML]{FEFEFF}0.4058 \\
RELP-Adap-1                 & \cellcolor[HTML]{FEFFFF}-0.3126 & \cellcolor[HTML]{BDD7EE}0.7404  & \cellcolor[HTML]{BDD7EE}1.1384 & \cellcolor[HTML]{BFD8EF}1.0097  & \cellcolor[HTML]{C1DAF0}0.6103 & \cellcolor[HTML]{C3DBF0}3.8314 \\
RELP-Adap-2                 & \cellcolor[HTML]{F8FBFE}-0.2408 & \cellcolor[HTML]{CFE2F3}0.5132  & \cellcolor[HTML]{D3E5F4}0.7969 & \cellcolor[HTML]{BDD7EE}1.0373  & \cellcolor[HTML]{BDD7EE}0.6422 & \cellcolor[HTML]{C3DBF0}3.8341 \\
\bottomrule
\end{tabular}
\end{table}

\subsection{Application to new data sets} \label{subsec:new_data}
In this section, we apply OLPS strategies---including our RELP-Adap-1 and RELP-Adap-2 strategies---to three new data sets, and compare their performance. In many OLPS studies, model parameters are tuned such that the strategies used perform well in the six (or some other number of) benchmark data sets. We focus on the case when $\gamma=0.2\%$ and the results when $\gamma=0.5\%$ are similar (see \ref{appdx:new_data_set}).

\begin{figure}[t] 
\centering
\begin{subfigure}{0.475\textwidth}
\centering
\includegraphics[scale=0.65]{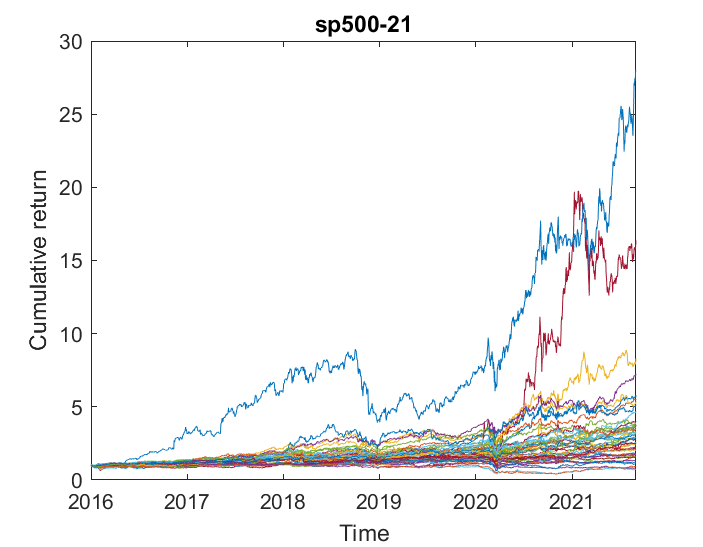}
\caption{Stock performance (initialized as $1$)} \label{fig:sp500-21_stock}
\end{subfigure}
\hfill
\begin{subfigure}{0.475\textwidth}
\centering
\includegraphics[scale=0.65]{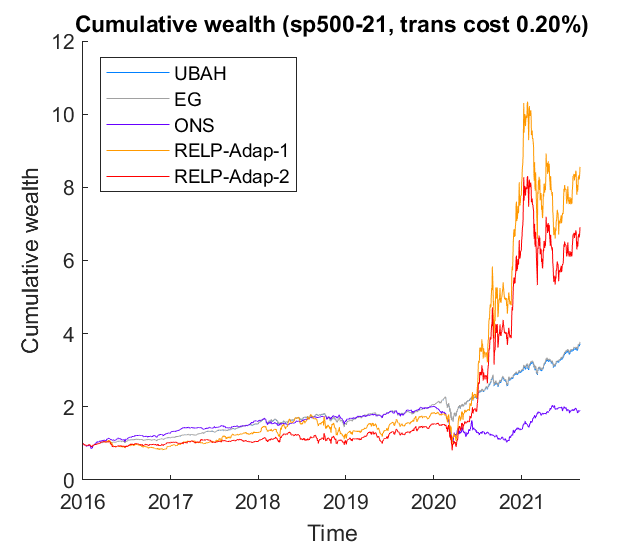}
\caption{Cumulative wealth of strategies} \label{fig:sp500-21_wealth}
\end{subfigure}
\caption{Performance of several OLPS strategies in SP500-21 (Top 3 stocks in order: TSLA, NVDA, PYPL)} \label{fig:sp500-21}
\end{figure}

Table~\ref{table:new_data_comparison} summarizes the cumulative wealth, Sharpe ratios and information ratios of the selected strategies. In terms of cumulative wealth, our RELP strategies outperform the other strategies when applied to the SP500-21 and HSI data sets. In particular, when applied to the SP500-21 data set, our RELP strategies generate almost twice as much wealth than that generated by the market strategy (i.e., UBAH). When applied to the CSI data set, the RELP-Adap-1 strategy yields satisfactory wealth that is close to the market strategy. Figure~\ref{fig:sp500-21_stock} provides a deeper understanding of our RELP strategies by giving the individual stock performances, whereas Figure~\ref{fig:sp500-21_wealth} shows the cumulative wealth of the strategies over time. Our RELP strategies rapidly recognize an uptrend, as they use a wide range of experts with different parameter choices. If one of these experts (e.g., a less risk-averse expert) captures this uptrend, our RELP strategies can yield superior cumulative wealth, as we monitor experts with outstanding returns. 

As illustrated in Figure~\ref{fig:sp500-21_wealth}, the design of adaptive schemes enables our strategies to adopt less risk-averse parameters than other strategies, and thereby generate better returns than other strategies. That is, the portfolio volatility associated with our strategies is larger than that associated with other strategies. However, the Sharpe ratios from the RELP strategies are competitive with those obtained from other strategies. The results for the information ratio, which is a measure used for comparing one strategy with the market strategy (as a benchmark), show that information ratios for the RELP strategies applied to the SP500-21 and HIS data sets are both positive. This means that our strategies have a competitive edge over the market strategy. 
    
Investors may change the input range of parameter values to better reflect their risk appetites. Thus, risk-averse investors may select a range of values that generate a conservative (low-volatility) investment. For instance, $\kappa\in[1,10]$ ensures there is a sufficiently robust selection-decision made (see \ref{appdx:sensitivity}). Moreover, there is flexibility in the selection of input parameters: rather than selecting values from a log-scale (which may result in experts having similar performances), we can select parameters that represent different risk levels (e.g., low, medium and high). Such an adaptive portfolio strategy is more intuitive for investors to apply.


%
\begin{table}[t]\small
\centering
\ra{0.6}
\caption{Comparison of performance of strategies applied to new data sets in terms of cumulative wealth (CW), Sharpe ratios (SR) and information ratio (IR)} \label{table:new_data_comparison}
\begin{tabular}{@{}rrrr|rrr|rrr@{}} \toprule
Strategy         & \multicolumn{3}{c}{SP500-21}                                                                       & \multicolumn{3}{c}{HSI}                                                                             & \multicolumn{3}{c}{CSI}                                                                            \\
($\gamma=0.2\%$) & CW                             & SR                              & IR                              & CW                              & SR                              & IR                              & CW                             & SR                              & IR                              \\ \cmidrule{2-10}
EG               & \cellcolor[HTML]{E3EEF8}3.7716 & \cellcolor[HTML]{BDD7EE}0.0669  & \cellcolor[HTML]{BDD7EE}0.0416  & \cellcolor[HTML]{D2E4F4}8.3710  & \cellcolor[HTML]{BDD7EE}0.0516  & \cellcolor[HTML]{BDD7EE}0.0453  & \cellcolor[HTML]{BED8EF}2.7275 & \cellcolor[HTML]{BED8EF}0.0261  & \cellcolor[HTML]{BDD7EE}-0.0004 \\
OLMAR            & \cellcolor[HTML]{FFFFFF}0.1958 & \cellcolor[HTML]{F9FBFE}-0.0356 & \cellcolor[HTML]{F8FBFE}-0.0833 & \cellcolor[HTML]{FFFFFF}0.0047  & \cellcolor[HTML]{FFFFFF}-0.0796 & \cellcolor[HTML]{FFFFFF}-0.1367 & \cellcolor[HTML]{FFFFFF}0.0078 & \cellcolor[HTML]{F6FAFD}-0.0805 & \cellcolor[HTML]{F5F9FD}-0.1232 \\
ONS              & \cellcolor[HTML]{F2F7FC}1.8802 & \cellcolor[HTML]{D5E6F5}0.0258  & \cellcolor[HTML]{E0EDF7}-0.0326 & \cellcolor[HTML]{FAFCFE}1.0942  & \cellcolor[HTML]{D7E7F5}0.0009  & \cellcolor[HTML]{E1EDF8}-0.0527 & \cellcolor[HTML]{C8DEF1}2.2872 & \cellcolor[HTML]{C0D9EF}0.0221  & \cellcolor[HTML]{C3DBF0}-0.0127 \\
RMR              & \cellcolor[HTML]{FFFFFF}0.1332 & \cellcolor[HTML]{FFFFFF}-0.0473 & \cellcolor[HTML]{FFFFFF}-0.0996 & \cellcolor[HTML]{FFFFFF}0.0053  & \cellcolor[HTML]{FFFFFF}-0.0801 & \cellcolor[HTML]{FFFFFF}-0.1344 & \cellcolor[HTML]{FFFFFF}0.0030 & \cellcolor[HTML]{FFFFFF}-0.0979 & \cellcolor[HTML]{FFFFFF}-0.1469 \\
TCO1             & \cellcolor[HTML]{F5F9FD}1.5364 & \cellcolor[HTML]{DAE9F6}0.0174  & \cellcolor[HTML]{E1EDF8}-0.0344 & \cellcolor[HTML]{F7FBFD}1.4953  & \cellcolor[HTML]{D2E4F4}0.0104  & \cellcolor[HTML]{E0ECF7}-0.0495 & \cellcolor[HTML]{F1F7FC}0.5934 & \cellcolor[HTML]{D1E3F4}-0.0106 & \cellcolor[HTML]{D4E5F4}-0.0495 \\
TCO2             & \cellcolor[HTML]{E7F1F9}3.2879 & \cellcolor[HTML]{CCE0F2}0.0417  & \cellcolor[HTML]{CFE2F3}0.0048  & \cellcolor[HTML]{EDF4FB}3.4832  & \cellcolor[HTML]{C9DFF2}0.0281  & \cellcolor[HTML]{D4E5F4}-0.0173 & \cellcolor[HTML]{E1EDF8}1.2570 & \cellcolor[HTML]{C8DEF1}0.0073  & \cellcolor[HTML]{C5DCF0}-0.0172 \\
UBAH             & \cellcolor[HTML]{E4EFF8}3.7260 & \cellcolor[HTML]{BED8EF}0.0668  & \cellcolor[HTML]{FFFFFF}NaN     & \cellcolor[HTML]{D4E5F4}7.9154  & \cellcolor[HTML]{BED8EF}0.0513  & NaN                             & \cellcolor[HTML]{BDD7EE}2.7323 & \cellcolor[HTML]{BDD7EE}0.0262  & NaN                             \\
UCRP             & \cellcolor[HTML]{E8F1FA}3.1192 & \cellcolor[HTML]{C1D9EF}0.0614  & \cellcolor[HTML]{E6F0F9}-0.0451 & \cellcolor[HTML]{F1F7FC}2.6686  & \cellcolor[HTML]{C9DFF2}0.0282  & \cellcolor[HTML]{E2EEF8}-0.0555 & \cellcolor[HTML]{C1DAEF}2.5911 & \cellcolor[HTML]{BED8EF}0.0251  & \cellcolor[HTML]{C2DBF0}-0.0114 \\
RELP-Adap-1      & \cellcolor[HTML]{BDD7EE}8.6093 & \cellcolor[HTML]{C1DAEF}0.0605  & \cellcolor[HTML]{BED8EF}0.0395  & \cellcolor[HTML]{C5DCF0}10.7291 & \cellcolor[HTML]{C0D9EF}0.0468  & \cellcolor[HTML]{C9DEF1}0.0144  & \cellcolor[HTML]{CDE1F3}2.0788 & \cellcolor[HTML]{C2DAF0}0.0178  & \cellcolor[HTML]{BED8EF}-0.0023 \\
RELP-Adap-2      & \cellcolor[HTML]{CBE0F2}6.9099 & \cellcolor[HTML]{C4DBF0}0.0563  & \cellcolor[HTML]{C1DAF0}0.0331  & \cellcolor[HTML]{BDD7EE}12.1288 & \cellcolor[HTML]{BFD8EF}0.0488  & \cellcolor[HTML]{C8DEF1}0.0167  & \cellcolor[HTML]{F2F7FC}0.5751 & \cellcolor[HTML]{D3E5F4}-0.0145 & \cellcolor[HTML]{D4E5F4}-0.0497 \\
\bottomrule
\end{tabular}
\end{table}

Finally, we note that the novel strategies do not always perform well when applied to the new data sets. Indeed, many of them fail to beat the market strategy in terms of cumulative wealth generation. This illustrates the problem of calibrating parameters using benchmark data sets: although strategies using some calibrated parameters may generate outstanding performances when applied to benchmark data sets (e.g., the TCO1 and TCO2 strategies), these parameters may not lead to a satisfactory performance in practice. Thus, it may not be prudent to rely on default parameters, as each data set has its own characteristics, and a single set of parameters cannot reflect all of these characteristics. This highlights the importance of developing adaptive schemes that have less reliance on parameters than previously developed schemes. Moreover, we investigate the features of our strategies, which provides crucial information to investors that are interested in applying these strategies in real-life settings.

\section{Conclusion} \label{sec:conclusion}
In this paper, we study the OLPS problem, and develop a novel robust strategy that is based on an ellipsoidal set and considers transaction costs. Robust optimization is now more prevalent than it was in previous years due to its ability to protect against some unfavorable scenarios, and optimization tools have been developed for the efficient solution of robust models. In the OLPS problem, which uses historical data that may be noisy, it is natural for an investor to hedge against unfavorable situations; however, there is little mention of such a use of robustness in the OLPS literature. The ellipsoidal uncertainty set we use in our novel strategy takes into account recent covariance structure. We reformulate the resulting optimization model into a second-order conic program that can be solved efficiently. 

Moreover, we address a crucial issue in current strategies for solving the OLPS problem that has not received sufficient attention: these strategies' reliance on parameters calibrated on benchmark data sets. As we see in our numerical experiment, such strategies may not perform as well as claimed when they are applied to some recent data sets, i.e., they may not be able to beat the market strategy in terms of returns. This highlights the important role played by model parameters, and we therefore develop novel adaptive schemes to address this problem in our model. Specifically, we borrow the idea of selecting the best problem and thus develop a sequential parameter-selection scheme based on cumulative returns. We also compare the numerical performances of our schemes and various other heuristic schemes, which reveals that our schemes are superior in most cases due to their ability to recognize market uptrend and protect investors from market downtrend. The schemes are also sufficiently flexible that investors can incorporate their risk attitude toward investment by controlling the range of input parameters. Finally, in extensive experiments using benchmark data sets and recently collected data sets, we observe that our strategies outperform many existing strategies. In particular, our strategies generate superior cumulative wealth.



To conclude, we develop a novel robust OLPS strategy with adaptive schemes that are used to select model parameters. Numerical experiments reveal that this adaptive scheme approach is promising, and demonstrate the competitive edge that our strategy has over other OLPS strategies. Nevertheless, there remain several areas for exploration in future work. First, a distributionally robust OLPS strategy that hedges against the worst-case distribution should be considered, as it would be interesting and useful to determine if the application of distributional robustness can improve portfolio performance. Second, as many OLPS strategies require the selection of a predictor $\xbt_{t+1}$, it will be crucial to study how a predictor may affect actual performance. In particular, the performance of our strategy may be enhanced if we employ machine learning tools to produce a more accurate predictor. Third, given the importance of developing adaptive schemes for model parameters, these schemes should be further investigated and improved. Finally, it would be interesting to explore how the use of our adaptive scheme may facilitate sequential decision-making processes under various application settings. 

\clearpage
\appheading
\appendix

\section{Existing OLPS strategies} \label{appdx:OLPS_strategy}

\subsection{Descriptions} \label{appdx:OLPS_description}
In this section, we provide more details on the OLPS strategies from difference classes mentioned in \ref{sec:literature_review}. First, in the class OLPS strategies that are based on the return momentum, \cite{Cover:1991} proposes the universal portfolio (UP) strategy. The idea is to assume that we have a set of portfolios (e.g., uniformly generated on a simplex set) and their cumulative wealth is used to construct a weight over all the portfolios. The weighted sum of these portfolios form our next investment proportion. \cite{Helmbold_et_al:1998} propose the exponential gradient (EG) strategy, which maximizes the log return with relative entropy regularization term. The most recent return vector is used as a proxy of the next period return. \cite{Agarwal_et_al:2006} consider the online Newton step (ONS) strategy, which maximizes the sum of historical log-returns with a regularization term (e.g., $\ell_2$ regularization). Recently, \cite{Lai_et_al:2017} propose the peak price tracking (PPT) strategy, which exploits the recent maximum price within a certain window size. 

Second, for the class of OLPS strategies that are based on the belief of mean reversion, one early paper that implements such an idea is \cite{Borodin_et_al:2004}. They propose the anti-correlation (Anticor) strategy which makes use of the cross-correlation matrix of the assets to determine the portfolio proportion that transfers from one asset to another under-performing asset. \cite{Li_et_al:2012} propose the passive-aggressive mean reversion (PAMR) strategy. The idea is to passively keep the portfolio if its return is below a certain threshold while actively rebalance it if its return is above the threshold. \cite{Li_et_al:2011b} and \cite{Li_et_al:2013} propose the confidence weighted mean reversion (CWMR) strategy. They assume the portfolio weight follows a Gaussian distribution. The distribution (i.e., mean and covariance matrix) is updated sequentially based on the mean reversion principle via Kullback–Leibler divergence. \cite{Li_et_al:2015} propose the online moving average reversion (OLMAR) strategy. The model itself is similar to PAMR but the return predictor used in the model is based on the moving average reversion predictors. \cite{Guo_et_al:2021} extends the OLMAR strategy by introducing adaptive adjustment of the moving average reversion parameter and propose the adaptive OLMAR (AOLMA) strategy. 

Finally, for patching-matching based strategies, \cite{Gyorfi_et_al:2006} propose the use of some kernel function to measure similarity. In particular, with the use of a uniform kernel, the similarity is measured by the Euclidean distance. \cite{Gyorfi_et_al:2008} propose the use of the $k$-nearest neighborhood to select historical returns. \cite{Li_et_al:2011} makes use of the correlation information to identify similar historical paths and propose the correlation-driven nonparametric learning (CORN) strategy. The idea is that what matters more would be the shape of the path instead of the numerical values (i.e., the returns themselves). 

\subsection{OLPS parameters} \label{appdx:OLPS_para}

In our numerical experiment (Section~\ref{sec:numerics}), we compare our proposed RELP strategy with several existing OLPS strategies. Specific parameter values in the strategies are determined as suggested in the corresponding paper. Below is the list of strategies for comparison.
\begin{enumerate}[label=(\alph*)]
\item UBAH: Market strategy
\item UCRP: Uniform constant rebalanced portfolio  
\item EG: Exponential gradient strategy with $\eta=0.05$ \citep{Helmbold_et_al:1998} 
\item ONS: Online Newton step with $\eta=0$, $\beta=1$ and $\delta=0.125$  \citep{Agarwal_et_al:2006} 
\item Anticor: The buy-and-hold Anticor-anticor strategy with $W=30$ \citep{Borodin_et_al:2004} 
\item CORN-U: Correlation-drive nonparametric learning strategy with uniform weights \citep{Li_et_al:2011b}
\item CORN-K: Correlation-drive nonparametric learning strategy with top-5 performing portfolios \citep{Li_et_al:2011b} 
\item PAMR: Passive-aggressive mean reversion strategy with $\epsilon=0.5$ \citep{Li_et_al:2012} 
\item CWMR-Var: Confidence weighted mean reversion strategy  with $\epsilon=0.5$ and $\phi=2$ \citep{Li_et_al:2011} 
\item OLMAR: Online moving average reversion strategy with $w=5$ and $\epsilon=10$ \citep{Li_et_al:2015} 
\item RMR: Robust mean reversion strategy with $w=5$ and $\epsilon=5$ \citep{Huang_et_al:2016} 
\item TCO1: Transaction cost optimization with $1/\xb_t$ predictor, $\eta=10$ and $\lambda=10\eta\gamma$ \citep{Li_et_al:2018} 
\item TCO2: Transaction cost optimization with moving average reversion predictor, $w=5$, $\eta=10$ and $\lambda=10\eta\gamma$ \citep{Li_et_al:2018}
\item AOLMA: Adaptive OLMAR strategy with $\epsilon=10$ and $\gamma=1e$-$5$ \citep{Guo_et_al:2021} 
\end{enumerate}

\section{Proofs in Section~\ref{sec:robust_model}} \label{appdx:proofs}

\subsection{Proof of Theorem~\ref{thm:TC_convex}}

\begin{proof}{Proof.}
For notational simplicity, we suppress the subscript for $\bbh_{t-1}$ and $\xbt_{t}$. Consider the following problem
\begin{equation} \label{eqn:TC_inter}
\begin{aligned}
& \underset{\bb}{\textup{maximize}}
& & \bb^\tp \xbt-\lambda\norms{\bbh-\bb}_1\\
& \textup{subject to}
& & \bb^\tp\one + \gamma\norms{\bbh-\bb}_1 = 1, \\
& & & \bb \geq 0.
\end{aligned}
\end{equation}
We first show that \eqref{eqn:TC} and \eqref{eqn:TC_inter} have the same optimal value and we can obtain an optimal solution for \eqref{eqn:TC} from that of \eqref{eqn:TC_inter}. 

Let $\bb_{(3)}^*$ be an optimal solution to \eqref{eqn:TC_inter}. Note that $\bb_{(3)}^*\ne\zero$ since $\zero$ is not feasible when $\gamma<1$. Define $\wt=\one^\tp\bb_{(3)}^* > 0$ and $\bbt=\bb_{(3)}^*/\wt \geq 0$. We claim that $(\wt,\bbt)$ is feasible to \eqref{eqn:TC}. First, we have
$$\wt+\gamma\norms{\bbh-\wt\bbt}_1=\one^\tp\bb_{(3)}^*+\gamma\norms{\bbh-\bb_{(3)}^*}_1=1,$$
where the last equality follows from the feasibility of $\bb_{(3)}^*$. Also, we have
$$\one^\tp \bbt=\one^\tp\bb_{(3)}^*/\wt=1.$$
This shows the feasibility of $(\wt,\bbt)$. Moreover, $(\wt,\bbt)$ is an optimal solution to \eqref{eqn:TC}. Suppose, on the contrary, that there exists another feasible solution $(\wbar,\bbbar)$ to \eqref{eqn:TC} which gives a better objective. Then, consider the point $\bb_{(3)}=\wbar\bbbar$. It is easy to see that $\bb_{(3)}$ is feasible to \eqref{eqn:TC_inter}. However, we have
$$\bb_{(3)}^\tp\xbt -\lambda\norms{\bbh-\bb_{(3)}}_1=\wbar\bbbar^\tp\xbt-\lambda\norms{\bbh-\wbar\bbbar}_1>\wt\bbt^\tp\xbt-\lambda\norms{\bbh-\wt\bbt}_1=(\bb_{(3)}^*)^\tp\xbt-\lambda\norms{\bbh -\bb_{(3)}^*}_1,$$
which contradicts the optimality of $\bb_{(3)}^*$.

Next, we show that \eqref{eqn:TC_convex} and \eqref{eqn:TC_inter} have the same optimal value and any optimal solution of \eqref{eqn:TC_convex} is an optimal solution to \eqref{eqn:TC_inter}. It is easy to see that the feasible set of \eqref{eqn:TC_inter} is a subset of that of \eqref{eqn:TC_convex}. It suffices to show that any optimal solution $\bb_{(2)}^*$ to \eqref{eqn:TC_convex} is feasible to \eqref{eqn:TC_inter}. That is, we need to show
$$\one^\tp\bb_{(2)}^*+\gamma\norms{\bbh-\bb_{(2)}^*}_1=1.$$
If $\lambda \geq \max_{i}\{\xt_{i}\}$, we will show that $\bbh$ is the unique optimal solution and the equality holds immediately. First, we claim that $\big(\bb_{(2)}^*\big)_i\geq\bbh_i$ for all $i\in\{1,\dots,m\}$. Suppose, on the contrary, that  $\big(\bb_{(2)}^*\big)_i < \bbh_i$ for some $i\in\{1,\dots,m\}$. Then, we have
\begin{align*}
\big(\bb_{(2)}^*\big)^\tp \xbt - \lambda \norms{\bbh-\bb_{(2)}^*}_1 &= \bbh^\tp \xbt + \sum_{i\in\calI^+}(\xt_i-\lambda)\Delta_i + \sum_{i\in \calI^-} (\xt_i + \lambda)\Delta_i \\
&< \bbh^\tp \xbt,
\end{align*}
where $\Delta_i = \big(\bb_{(2)}^*\big)_i - \bbh_i $ with $\calI^+=\{i\mid \Delta_i > 0\}$ and $\calI^- =\{i\mid \Delta_i <0\}\ne\emptyset$. This contradicts that $\bb_{(2)}^*$ is optimal. Finally, since $\bbh^\tp\one = 1$, we must have $\bb_{(2)}^*=\bbh$ as the unique solution.

Suppose that $\lambda < \max_{i}\{\xt_{i}\}$. To show the desired equality, we claim that there exists $i''\in\argmax_i \{\xt_i\}:=\calI_\text{max}$ such that $\big(b_{(2)}^*\big)_{i''} \geq \bh_{i''}$. First, note that there exists $i'\in\{1,\dots,m\}$ such that $\big(b^*_{(2)}\big)_{i'} \geq \bh_{i'}$ (otherwise, $\bbh$ is a feasible solution with a better objective). Suppose, on the contrary, that $\big(b^*_{(2)}\big)_{i''} <\bh_{i''}$ for all $i''\in\argmax\{\xt_i\}$. 
\begin{enumerate}
\item[(i)]  If $\big(b^*_{(2)}\big)_{i'} > \bh_{i'}$, define $\pi=\min\big\{\big(b^*_{(2)}\big)_{i'}-\bh_{i'},\,\bh_{i''}-\big(b^*_{(2)}\big)_{i''}\big\}>0$ for some $i''\in\calI_\text{max}$ and consider the vector $\bb_{(2)}$ defined by $$\big(b_{(2)}\big)_i=\begin{cases} \big(b_{(2)}^*\big)_{i}+\pi & \text{if } i=i'', \\ \big(b_{(2)}^*\big)_{i}-\pi & \text{if } i=i', \\ \big(b_{(2)}^*\big)_{i}  & \text{otherwise. } \end{cases}$$ It is easy to see that 
\begin{align*}
\big| \big(b_{(2)}\big)_{i'}-\bh_{i'} \big| &= \big| \big(b^*_{(2)}\big)_{i'}-\bh_{i'} \big| - \pi, \\
\big| \big(b_{(2)}\big)_{i''}-\bh_{i''} \big| &= \big| \big(b^*_{(2)}\big)_{i''}-\bh_{i''} \big| - \pi.
\end{align*}
Hence, $\bb_{(2)}$ is feasible to \eqref{eqn:TC_convex} with objective value
\begin{align*}
\bb_{(2)}^\tp\xbt-\lambda\norms{\bb_{(2)}-\bbh}_1 &= \big(\bb^*_{(2)}\big)^\tp\xbt-\lambda\norms{\bb^*_{(2)}-\bbh}_1 + \pi \xt_{i''} + \pi\lambda - \pi\xt_{i'}+\pi\lambda \\
&=  \big(\bb^*_{(2)}\big)^\tp\xbt-\lambda\norms{\bb^*_{(2)}-\bbh}_1 + \pi (\xt_{i''}- \xt_{i'}) + 2\pi\lambda\\
&> \big(\bb^*_{(2)}\big)^\tp\xbt-\lambda\norms{\bb^*_{(2)}-\bbh}_1,
\end{align*}
which contradicts the optimality of $\bb^*_{(2)}$.

\item[(ii)] If $\big(b^*_{(2)}\big)_{i'} = \bh_{i'}$, define $\pi= \bh_{i''}-\big(b^*_{(2)}\big)_{i''} >0$ for some $i''\in\calI_\text{max}$ and $\bb_{(2)}$ accordingly as in (i). It is easy to check
\begin{align*}
\big| \big(b_{(2)}\big)_{i'}-\bh_{i'} \big| &= \big| \big(b^*_{(2)}\big)_{i'}-\bh_{i'} \big| + \pi, \\
\big| \big(b_{(2)}\big)_{i''}-\bh_{i''} \big| &= \big| \big(b^*_{(2)}\big)_{i''}-\bh_{i''} \big| - \pi.
\end{align*}
Hence, $\bb_{(2)}$ is feasible to \eqref{eqn:TC_convex} with objective value
\begin{align*}
\bb_{(2)}^\tp\xbt-\lambda\norms{\bb_{(2)}-\bbh}_1 &= \big(\bb^*_{(2)}\big)^\tp\xbt-\lambda\norms{\bb^*_{(2)}-\bbh}_1 + \pi \xt_{i''} + \pi\lambda - \pi\xt_{i'}-\pi\lambda \\
&=  \big(\bb^*_{(2)}\big)^\tp\xbt-\lambda\norms{\bb^*_{(2)}-\bbh}_1 + \pi (\xt_{i''}- \xt_{i'})\\
&> \big(\bb^*_{(2)}\big)^\tp\xbt-\lambda\norms{\bb^*_{(2)}-\bbh}_1,
\end{align*}
which contradicts the optimality of $\bb^*_{(2)}$.
\end{enumerate}
The above two cases show that there exists $i''\in\argmax_i \{\xt_i\}:=\calI_\text{max}$ such that $\big(b_{(2)}^*\big)_{i''} \geq \bh_{i''}$. To proceed, we fix a particular choice $i''\in\calI_\text{max}$.

Finally, we prove that the equality holds by considering following two cases.
\begin{enumerate}
\item[(a)] If $\big(b_{(2)}^*\big)_{i''}=1$, since $\one^\tp \bb_{(2)}^* \leq 1$ is implied from the constraint, we must have $\bb_{(2)}^*=e_{i''}$, i.e., a vector of zeros except the $i''$th entry being $1$. If $\bbh=e_{i''}$, then we obviously have  $\one^\tp\bb_{(2)}^*+\gamma\norms{\bbh-\bb_{(2)}^*}_1=1$. If $\bbh\ne e_{i''}$, this would lead to a contradiction since $$\one^\tp\bb_{(2)}^*+\gamma\norms{\bbh-\bb_{(2)}^*}_1=1+\gamma\norms{\bbh-e_{i''}}_1>1.$$

\item[(b)] Suppose that $\big(b_{(2)}^*\big)_{i''}<1$. If $\one^\tp\bb_{(2)}^*+\gamma\norms{\bbh-\bb_{(2)}^*}_1=1$, then the proof is completed. Otherwise, consider that $\one^\tp\bb_{(2)}^*+\gamma\norms{\bbh-\bb_{(2)}^*}_1=\delta<1$. Note that this implies $\bb_{(2)}^*\ne\bbh$ and $\one^\tp\bb_{(2)}^*< 1$ since $\bbh^\tp\one = 1$. Define $\theta=(1-\delta)/(1+\gamma)>0$. Consider the vector $\bb_{(2)}$ defined as $$\big(b_{(2)}\big)_i=\begin{cases} \big(b_{(2)}^*\big)_{i''}+\theta & \text{if } i=i'', \\  \big(b_{(2)}^*\big)_{i}  & \text{if } i\ne i''.\end{cases}$$ By construction, $\bb_{(2)}$ is feasible to \eqref{eqn:TC_convex}. Indeed,
\begin{align*}
\one^\tp\bb_{(2)}+\gamma\norms{\bbh-\bb_{(2)}}_1  &=\one^\tp\bb_{(2)}^*+\theta + \gamma \sum_{i=1,i\ne i''}^m \Big|\bh_i-\big(b_{(2)}\big)_i \Big| + \gamma\Big|\bh_{i''}-\big(b_{(2)}\big)_{i''} \Big| \\
&= \one^\tp\bb_{(2)}^*+\theta + \gamma \sum_{i=1,i\ne i''}^m \Big|\bh_i-\big(b_{(2)}^*\big)_i \Big| + \gamma\bigg\{\Big|\bh_{i''}-\big(b_{(2)}^*\big)_{i''} \Big| + \theta \bigg\}\\
&= \one^\tp\bb_{(2)}^* + \gamma\norms{\bbh-\bb_{(2)}^*}_1 + \theta(1+\gamma) \\
&= \delta + (1-\delta) \\
&= 1.
\end{align*}
However, from the assumption that $\lambda < \max_i\{\tilde{x}_i\} = \xt_{i''}$, we have
$$\xbt^\tp\bb_{(2)}-\lambda\norms{\bbh-\bb_{(2)}}_1 =\xbt^\tp\bb_{(2)}^*-\lambda\norms{\bbh-\bb_{(2)}^*}_1 + \theta\tilde{x}_{i''}-\theta\lambda >\xbt^\tp\bb_{(2)}^*-\lambda\norms{\bbh-\bb_{(2)}^*}_1, $$
which contradicts the optimality of $\bb_{(2)}^*$. 
\end{enumerate}
Concluding from these two cases, we show that an optimal solution of \eqref{eqn:TC_convex} is always feasible to \eqref{eqn:TC_inter}. This completes the proof. 
\end{proof}

\subsection{Proof of Corollary~\ref{coro:TC_LP}}

\begin{proof}{Proof.}
Let $(\bb^*,\ub^*,\vb^*)$ be an optimal solution to \eqref{eqn:TC_LP}. It is straightforward to show that $\ub^*=\max\{\bbh_{t-1}-\bb^*,0\}$ and $\vb^*=\max\{\bb^*-\bbh_{t-1},0\}$. Hence, $(\ub^*+\vb^*)^\tp\one=\norms{\bbh_{t-1}-\bb^*}_1$ and it is easy to see that $\bb^*$ is an optimal solution to \eqref{eqn:TC_convex}. Thus, the equivalence follows from Theorem~\ref{thm:TC_convex}. 
\end{proof}

\subsection{Proof of Proposition~\ref{prop:inner_max}}

\begin{proof}{Proof.}
We adopt the idea of proof from \cite{Feng_Palomar:2016}. Note that
\begin{align*}
\calU(\xbt_t,\kappa) &= \Big\{\xb\in\R^m \, \Big| \,(\xb-\xbt_t)^\tp\Sigmab^{-1}(\xb-\xbt_t) \leq \kappa^2\Big\}\\
&=\Big\{\xb\in\R^m \, \Big| \,\yb^\tp\Ub^{-1}\Ub^{-\top}\yb\leq \kappa^2,\, \yb=\xb-\xbt_t \Big\} \\
&=\Big\{\xb\in\R^m \, \Big| \,\norms{\Ub^{-\top}\yb}_2\leq \kappa,\, \xb=\yb+\xbt_t \Big\}\\
&=\Big\{\xb\in\R^m \, \Big| \,\norms{\ybt}_2\leq \kappa,\, \xb=\Ub^\tp\ybt+\xbt_t \Big\}\\
&=\Big\{\xb\in\R^m \, \Big| \,\norms{\ybt}_2\leq 1,\, \xb=\kappa\Ub^\tp\ybt+\xbt_t \Big\}.
\end{align*}
Therefore,
\begin{align*}
\min_{\xb\in\calU(\xbt_t,\kappa)} \big\{w\bb^\tp \xb \big\} &= \xbt_t^\tp\bb + \kappa \min_{\ybt:\, \norms{\ybt}_2 \leq 1} \bb^\tp\Ub^\tp\ybt = \xbt_t^\tp\bb + \kappa \min_{\ybt:\, \norms{\ybt}_2 \leq 1} (\Ub\bb)^\tp\ybt \\
&=   \xbt_t^\tp\bb - \kappa \norms{\Ub\bb}_2.
\end{align*}
This completes the proof. 
\end{proof}

\subsection{Proof of Lemma~\ref{lem:Lip_l2_norm}}

\begin{proof}{Proof.}
With the use of reverse triangle inequality, we have
$$\Big|\norms{\Ub\bb_1}_2- \norms{\Ub\bb_2}_2\Big| \leq \norm{\Ub(\bb_1-\bb_2)}_2 \leq \norms{\Ub}_F \norms{\bb_1-\bb_2}_2,$$
where $\norms{\cdot}_F$ denotes the Frobenius norm. If we write $\Thetab=\diag(\sigma_1,\dots,\sigma_m)$ and $\Cb=\Ub_c^\tp \Ub_c$ as the Cholesky factorization of the correlation matrix, we have
$$\Ub^\tp\Ub = \Sigmab = \Thetab\big(\Ub_c^\tp \Ub_c\big)\Thetab.$$
Note that the squared sum of the entries in a column of $\Ub_c$ is always $1$ since $\Cb$ is a correlation matrix (see \citealp{Madar:2015}). As $\Ub = \Ub_c\Thetab$, the $i$th column of the $\Ub$ equals the $i$th column of $\Ub_c$ multiplied by $\sigma_i$ for $i\in\{1,\dots,m\}$. Hence, we can deduce that $\norms{\Ub}_F=\sigma$ and this completes the proof. 
\end{proof}

\subsection{Proof of Theorem~\ref{thm:TC_robust_SOCP}}

\begin{proof}{Proof.}
Again, we suppress the subscript on $\bbh_{t-1}$ and $\xbt_t$ for notational simplicity. First, consider the following problem
\begin{equation} \label{eqn:TC_robust_inter}
\begin{aligned}
& \underset{\bb}{\textup{maximize}}
& & \bb^\tp_t \xbt-\lambda\norms{\bbh-\bb}_1-\kappa\norms{\Ub\bb}_2\\
& \textup{subject to}
& & \bb^\tp\one + \gamma\norms{\bbh-\bb}_1 = 1, \\
& & & \bb \geq 0.
\end{aligned}
\end{equation}
In view of Proposition~\ref{prop:inner_max}, the same argument in the first part of the proof of Theorem~\ref{thm:TC_convex} shows that if $\bb^*_{(3)}$ is an optimal solution to \eqref{eqn:TC_robust_inter}, then $(\wt,\bbt):=(\one^\tp\bb_{(3)}^*, \bb_{(3)}^*/\wt)$ is an optimal solution to \eqref{eqn:TC_robust}. 

Finally, we need to show that an optimal solution $\bb_{(2)}^*$ of \eqref{eqn:TC_robust_SOCP} achieves equality in the first constraint. Suppose, on the contrary, that $\big(\bb^*_{(2)}\big)^\tp\one + \gamma\norms{\bbh-\bb^*_{(2)}}_1 = \delta < 1$. Fix $i'\in\argmax_i\{\xt_i\}$. If $\big(b^*_{(2)}\big)_{i'}=1$, then the argument (a) in the proof of Theorem~\ref{thm:TC_convex} applies. That is, if $\bbh=e_{i'}$, then the equality holds; otherwise, contradiction is reached. If $\big(b^*_{(2)}\big)_{i'}<1$, we consider the following two cases.
\begin{enumerate}
\item[(1)] Assume that $\big(b^*_{(2)}\big)_{i'} \geq \bh_{i'}$. Define $\theta = (1-\delta)/(1+\gamma)>0$ and $\bb_{(2)}$ by $$\big(b_{(2)}\big)_i=\begin{cases} \big(b_{(2)}^*\big)_{i}+\theta & \text{if } i=i', \\ \big(b_{(2)}^*\big)_{i}  & \text{otherwise. } \end{cases}$$ By construction, $\bb_{(2)}$ is feasible to \eqref{eqn:TC_robust_SOCP} but
\begin{align*}
&\quad\, \Big[\xbt^\tp\bb_{(2)} - \lambda \norms{\bbh-\bb_{(2)}}_1 - \kappa\norms{\Ub\bb_{(2)}}_2 \Big] - \Big[\xbt^\tp\bb^*_{(2)} - \lambda \norms{\bbh-\bb^*_{(2)}}_1 - \kappa\norms{\Ub\bb^*_{(2)}}_2 \Big] \\
&=\theta \xt_{i'}-\theta \lambda + \kappa\Big[\norms{\Ub\bb^*_{(2)}}_2-\norms{\Ub\bb_{(2)}}_2\Big] \\
&> \theta\xt_{i'} - \theta\lambda-\kappa\sigma\theta \\
&= \theta(\xt_{i'}-\lambda-\kappa\sigma) \\
&>0,
\end{align*}
where the first inequality follows from Lemma~\ref{lem:Lip_l2_norm} and the last inequality follows from the assumption that $\max_i\{\xt_i\}=\xt_{i'}>\kappa\sigma + \lambda$. This contradicts the optimality of $\bb_{(2)}^*$.

\item[(2)] Assume that $\big(b^*_{(2)}\big)_{i'} < \bh_{i'}$. Define $\theta =\min\{ (1-\delta)/(1-\gamma),\, \bh_{i'}-\big(b^*_{(2)}\big)_i\}>0$ and $\bb_{(2)}$ as in (1). By construction, $\bb_{(2)}$ is feasible to \eqref{eqn:TC_robust_SOCP}. Indeed,
\begin{align*}
\bb_{(2)}^\tp\one + \gamma\norms{\bbh-\bb_{(2)}}_1 &= \big(\bb^*_{(2)}\big)^\tp\one + \theta + \gamma\norms{\bb^*_{(2)}-\bbh}_1 -\gamma\theta \\
&= \delta + (1-\gamma)\theta \\
&\leq 1.
\end{align*}
However, we have
\begin{align*}
&\quad\, \Big[\xbt^\tp\bb_{(2)} - \lambda \norms{\bbh-\bb_{(2)}}_1 - \kappa\norms{\Ub\bb_{(2)}}_2 \Big] - \Big[\xbt^\tp\bb^*_{(2)} - \lambda \norms{\bbh-\bb^*_{(2)}}_1 - \kappa\norms{\Ub\bb^*_{(2)}}_2 \Big] \\
&=\theta \xt_{i'}+\theta \lambda + \kappa\Big[\norms{\Ub\bb^*_{(2)}}_2-\norms{\Ub\bb_{(2)}}_2\Big] \\
&> \theta\xt_{i'} + \theta\lambda-\kappa\sigma\theta \\
&= \theta(\xt_{i'}+\lambda-\kappa\sigma) \\
&>0,
\end{align*}
where the first inequality follows from Lemma~\ref{lem:Lip_l2_norm} and the last inequality follows from the assumption that $\max_i\{\xt_i\}=\xt_{i'}>\kappa\sigma + \lambda>\kappa\sigma - \lambda$. This contradicts the optimality of $\bb_{(2)}^*$.
\end{enumerate}
Summarizing the two cases, we thus have $\big(\bb^*_{(2)}\big)^\tp\one + \gamma\norms{\bbh-\bb^*_{(2)}}_1 = 1$. 
\end{proof}

\section{Adaptive schemes} \label{appdx:adaptive_scheme}
\subsection{The adaptive scheme on \texorpdfstring{$\lambda$}{TEXT}} \label{appdx:adaptive_lambda}

In this section, we provide the algorithm details of the proposed adaptive scheme on $\lambda$. However, as mentioned in Section~\ref{sec:adaptive_scheme}, this could also be applied on $\kappa$. Algorithm~\ref{algo:adaptive_lambda} summarizes the details of the adaptive scheme. It takes a range of values of $\lambda$ as input. The window size $W$ is related to the moving average we computed at each period. The indifference zone parameter $\delta$ could be set to a value greater than zero when the asset returns are stable over time to avoid active rebalancing. At each period, based on the most recent $W$ day cumulative wealth $\{S^l_{t-W},\dots,S^l_{t-1}\}$ of the current tracking expert $l'$, we compute the sample mean $\bar{S}^l_{t-1}$ and sample standard deviation $\hat{\sigma}^l_{t-1}$. If the cumulative wealth moving average of another expert portfolio, say $l''$, exceeds the upper limit $\bar{S}^l_{t-1}+1.96\hat{\sigma}^l_{t-1}$, as well as the indifference zone parameter $\delta$, we switch to the new expert in the following periods. Note that we do not have any distributional assumption on the return series, we construct the upper limit via the approximation of normal distribution. However, we remark that this could be adjust be the users. This controls the trade-off between exploration (switching to other experts) and exploitation (staying with the current expert), which is related to investor's risk preferences.
\LinesNumbered  
\IncMargin{1em}
\begin{algorithm}[t] 
\SetKwInOut{Initialization}{Initialization}

\Initialization{Set $\bb_1=\one/m$, $\bb_0=\zero$ and $S_0=1$. Pick a range of values of $\lambda$, i.e., $\{\lambda_1,\dots,\lambda_L\}$. Choose window size $W$, indifference zone parameter $\delta$ and the initial tracking portfolio $l'$.}
\For{$t=1,\dots,n$}{
Obtain the new portfolio weights $\bb_t^l$ for expert with $\lambda_l$ for $l\in\{1,\dots,L\}$.\\
(For $t=1$, we use $\bb^l_1=\one/m$; for $t\in\{2,\dots,m+1\}$, we use $\bb^l_t = \bbh^l_{t-1}$.) \\
Set $\bb_t=\bb_t^{l'}$. \\
Compute the net proportion $w_{t-1}$ of the rebalance (from $\bbh_{t-1}$), as well as $w^l_{t-1}$ for each expert (from $\bbh_{t-1}^k$) by solving \eqref{eqn:weight_balance}. \\
\BlankLine
Observe the realized return $\xb_t$. \\
Update the cumulative wealth as $S_t = S_{t-1} w_{t-1}(\bb_{t}^\tp \xb_{t})$. \\
Update the cumulative wealth for each expert as $S_t^l= S_{t-1}^l w_{t-1}^l[(\bb_{t}^l)^\tp \xb_{t}]$ for $k\in\{1,\dots,N\}$. \\
Based on the recent portfolio wealth $\{S^{l'}_{t-W+1},\dots,S^{l'}_{t}\}$ of expert $l'$, compute the sample mean $\bar{S}^l_{t}$ and sample standard deviation $\hat{\sigma}^{l'}_{t}$. \\
\If{$\exists l''\in\{1,\dots,L\}\setminus\{l'\}$ \textup{such that} $\bar{S}_t^{l''} >\bar{S}^{l'}_t+\max\{\delta, 1.96\hat{\sigma}^{l'}_t\}$ }{
Update the tracking expert $l'=l''$.
}
Obtain the current portfolio weights $\bbh_t = \bb_t\cdot \xb_t / \bb_t^\tp \xb_t$ as well as $\bbh_t^l$ for each expert. 
}
\BlankLine
\caption{An adaptive scheme for $\lambda$}\label{algo:adaptive_lambda}
\end{algorithm}\DecMargin{1em}

\subsection{The alternative adaptive scheme on \texorpdfstring{$\kappa$}{TEXT}} \label{appdx:adaptive_kappa}

Next, we provide the algorithm details of the proposed adaptive scheme on $\kappa$, summarized as Algorithm~\ref{algo:adaptive_kappa}. Similar to Algorithm~\ref{algo:adaptive_lambda}, it takes a possibly wide range of values of $\kappa$, say $\{\kappa_1,\dots,\kappa_N\}$, as input. At each period, we update the portfolio of each expert (corresponding to each choice of $\kappa$), as well as the corresponding cumulative wealth. Then, at period $t$, we have $N$ different cumulative wealth up to $t-1$ for the experts. By sorting the values of $\kappa$ in ascending order of the cumulative wealth levels, we set $\kappa_t$ as the average of the top-$K$ performing values of $\kappa$. 

\begin{rem}
In our numerical experiments, $\kappa$ is chosen in a range of log-scaled values and hence, we take the geometric average of the top-$K$ performing values of $\kappa$.
\end{rem}

%
\LinesNumbered  
\IncMargin{1em}
\begin{algorithm}[t] 
\SetKwInOut{Initialization}{Initialization}

\Initialization{Set $\bb_1=\one/m$, $\bb_0=\zero$ and $S_0=1$. Pick a range of values of $\kappa$, i.e., $\{\kappa_1,\dots,\kappa_N\}$. Choose $K$ and initial $\kappa_{m+2}$.}
\For{$t=1,\dots,n$}{
Obtain the new portfolio weights $\bb_t$ by solving \eqref{eqn:TC_robust_SOCP} with $\kappa=\kappa_{t-1}$. \\
(For $t=1$, we use $\bb_1=\one/m$; for $t\in\{2,\dots,m+1\}$, we use $\bb_t = \bbh_{t-1}$.) \\
Obtain the new portfolio weights $\bb_t^k$ for expert with $\kappa_k$ by solving \eqref{eqn:TC_robust_SOCP} for $k\in\{1,\dots,N\}$.\\
(For $t=1$, we use $\bb^k_1=\one/m$; for $t\in\{2,\dots,m+1\}$, we use $\bb^k_t = \bbh^k_{t-1}$.) \\
Compute the net proportion $w_{t-1}$ of the rebalance (from $\bbh_{t-1}$), as well as $w^k_{t-1}$ for each expert (from $\bbh_{t-1}^k$) by solving \eqref{eqn:weight_balance}. \\
\BlankLine
Observe the realized return $\xb_t$. \\
Update the cumulative wealth as $S_t = S_{t-1} w_{t-1}(\bb_{t}^\tp \xb_{t})$. \\
Update the cumulative wealth for each expert as $S_t^k= S_{t-1}^k w_{t-1}^k[(\bb_{t}^k)^\tp \xb_{t}]$ for $k\in\{1,\dots,N\}$. \\
Sort $S_t^k$ in ascending order.\\
Compute $\kappa_t$ as the average value of $\kappa$'s of the top-$K$ performing portfolios. \\
Obtain the current portfolio weights $\bbh_t = \bb_t\cdot \xb_t / \bb_t^\tp \xb_t$ as well as $\bbh_t^k$ for each expert. 
}
\BlankLine
\caption{An adaptive scheme for $\kappa$}\label{algo:adaptive_kappa}
\end{algorithm}\DecMargin{1em}

\section{New data sets details} \label{appdx:data_set}
\begin{figure}[t]
\centering
\begin{subfigure}{0.475\textwidth}
\centering
\includegraphics[scale=0.65]{fig_sp500-21_stocks.png}
\caption{SP500-21} \label{fig:sp500-21_data_set}
\end{subfigure}
\begin{subfigure}{0.475\textwidth}
\centering
\includegraphics[scale=0.65]{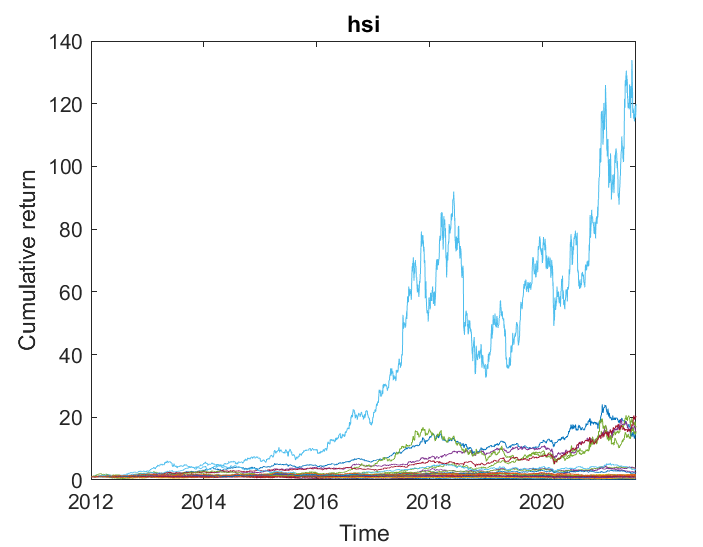}
\caption{HSI} \label{fig:hsi_data_set}
\end{subfigure} \\[1ex]
\begin{subfigure}{\textwidth}
\centering
\includegraphics[scale=0.65]{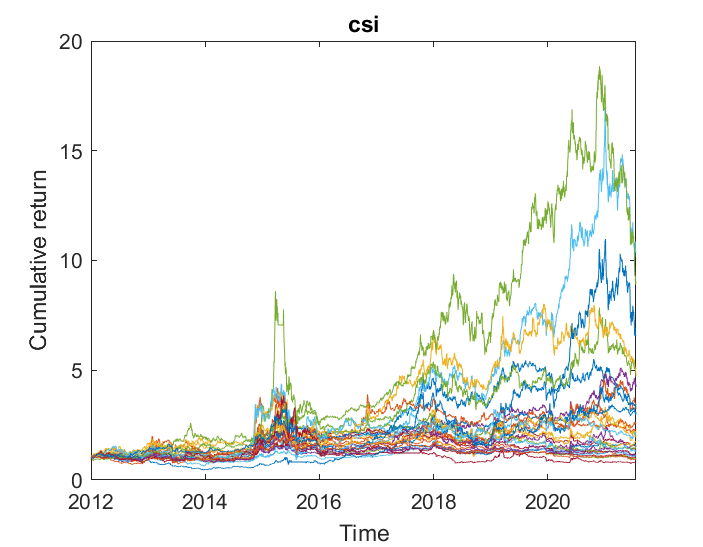}
\caption{CSI} \label{fig:csi_data_set}
\end{subfigure}
\caption{Stock performances in new data sets (initialized as $1$)} \label{fig:new_data_set}
\end{figure}

In this section, we provide some details of the new data sets. Table~\ref{table:stock_codes} shows the stock codes in each new data set: SP500-21, HSI and CSI. There are $50$ stocks in SP500-21 and the data set covers the period from Jan 2016 to Aug 2021. Figure~\ref{fig:sp500-21_data_set} plots the individual stock cumulative returns over the period. We observe that most of them experience a mild growth in return and a few of them has significantly higher returns than the other. For HSI and CSI, there are $30$ stocks and the data sets cover the period from Jan 2012 to Aug 2021. Figure~\ref{fig:hsi_data_set} shows the individual stock cumulative returns for HSI data set. During this period, most of them did not experience a significant growth in return. However, there is one stock with exceptionally high return and a few dominating stocks (with higher cumulative return). Finally, Figure~\ref{fig:csi_data_set} presents the individual stock cumulative returns for CSI data set. Most of the stocks could result in a growth in cumulative return but to a different extent.

\begin{table}[t]\small
\centering
\ra{0.6}
\caption{Stock codes in the new data sets (in alphabetical/ascending order)} \label{table:stock_codes}
\begin{tabular}{@{}cccccccccc@{}} \toprule
\multicolumn{10}{c}{SP500-21} \\ \midrule
AAPL & ABBV & ABT  & ACN & ADBE & AMZN & AVGO & BAC  & BRK-B  & CMCSA \\
COST & CRM  & CSCO & CVX & DHR  & DIS  & FB   & GOOG & GOOG-L & HD    \\
INTC & JNJ  & JPM  & KO  & LIN  & LLY  & MA   & MCD  & MDT    & MRK   \\
MSFT & NEE  & NFLX & NKE & NVDA & ORCL & PEP  & PFE  & PG     & PYPL  \\
T    & TMO  & TSLA & TXN & UNH  & V    & VZ   & WFC  & WMT    & XOM   \\ \midrule
\multicolumn{10}{c}{HSI} \\ \midrule
0001 & 0002 & 0003 & 0005 & 0011 & 0016 & 0027 & 0066 & 0175 & 0267 \\
0388 & 0669 & 0688 & 0700 & 0857 & 0883 & 0939 & 0941 & 1109 & 1299 \\
1398 & 1928 & 2020 & 2313 & 2318 & 2382 & 2388 & 2628 & 3328 & 3988 \\ \midrule
\multicolumn{10}{c}{CSI} \\ \midrule
\multicolumn{2}{c}{000002.SS} & \multicolumn{2}{c}{600000.SS}  & \multicolumn{2}{c}{600015.SS}  & \multicolumn{2}{c}{600016.SS}  & \multicolumn{2}{c}{600030.SS}\\
\multicolumn{2}{c}{600036.SS} & \multicolumn{2}{c}{600048.SS}  & \multicolumn{2}{c}{600104.SS}  & \multicolumn{2}{c}{600276.SS}  & \multicolumn{2}{c}{600267.SS}\\
\multicolumn{2}{c}{600837.SS} & \multicolumn{2}{c}{600887.SS}  & \multicolumn{2}{c}{600900.SS}  & \multicolumn{2}{c}{601166.SS}  & \multicolumn{2}{c}{601169.SS}\\
\multicolumn{2}{c}{601288.SS} & \multicolumn{2}{c}{601318.SS}  & \multicolumn{2}{c}{601328.SS}  & \multicolumn{2}{c}{601398.SS}  & \multicolumn{2}{c}{601601.SS}\\
\multicolumn{2}{c}{601668.SS} & \multicolumn{2}{c}{601688.SS}  & \multicolumn{2}{c}{601766.SS}  & \multicolumn{2}{c}{601818.SS}  & \multicolumn{2}{c}{601988.SS}\\
\multicolumn{2}{c}{601989.SS} & \multicolumn{2}{c}{000001.SZ}  & \multicolumn{2}{c}{000651.SZ}  & \multicolumn{2}{c}{000725.SZ}  & \multicolumn{2}{c}{000858.SZ}\\
\bottomrule
\end{tabular}
\end{table}

\section{Additional computational results} \label{appdx:add_comp_results}

\subsection{Performance metric} \label{appdx:performance_metric}

In Section~\ref{sec:numerics}, we examine the OLPS strategies based on several performance metrics. We provide a brief description of the metrics and the associated mathematical formula are readily available in the literature.
\begin{enumerate}
\item Cumulative wealth: Terminal wealth with initial wealth being $1$ 
\item Mean excess return (MER): Difference in the average daily return of the strategy and the market portfolio \citep{Jegadeesh:1990} 
\item Sharpe ratio: Excess mean return divided by its standard deviation with annual risk-free rate $4\%$ (same rate as used in the OLPS package by \citealp{Li_et_al:2016}) 
\item Calmar ratio: Annualized return divided by maximum drawdown \citep{Young:1991} 
\item Information ratio: Mean difference in excess market return divided by its standard deviation \citep{Treynor_Black:1973}
\item Maximum drawdown: Maximum percentage loss of wealth from a past peak to a past trough \citep{Magdon-Ismail_Atiya:2004} 
\item Value-at-risk: The $\alpha$-quantile of the loss distribution \citep{Jorion:2001}
\end{enumerate}

\subsection{Sensitivity analysis} \label{appdx:sensitivity}

In this section, we demonstrate how the choice of parameters will affect the overall strategy performance. In particular, we focus on $\gamma=0.2\%$ with MSCI and NYSE-N data sets. The observations are similar for the other cases so we just briefly mention the other data sets at the end. We focus on the comparison of the cumulative wealth, Sharpe ratio, maximum drawdown and Calmar ratio for different combinations of $\kappa$ and $\lambda$. We choose $21$ values in a log-scale of $\kappa$ between $0.1$ and $10$ inclusive while we pick $\lambda=\delta\gamma$ with $\delta$ taking $41$ values in a log-scale between $0.01$ and $100$ inclusive. The ranges cover a wide extent of robustness and trade off between return and transaction costs.

\begin{figure}[p]
\includegraphics[scale=0.7]{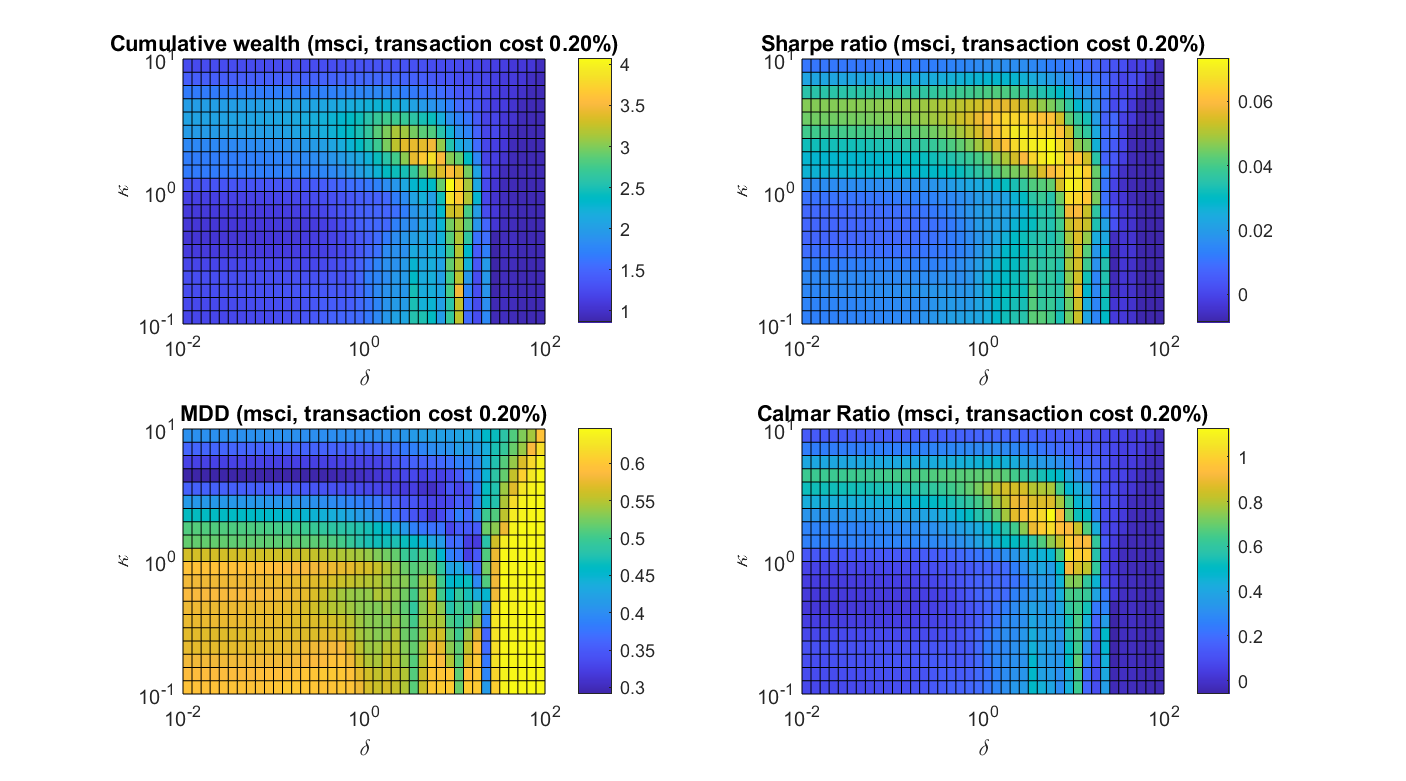}
\caption{Sensitivity on parameters on MSCI data set with $\gamma=0.2\%$} \label{fig:sensitivity_msci_gamma_0.002}
\end{figure}

Figure~\ref{fig:sensitivity_msci_gamma_0.002} shows the sensitivity analysis on MSCI data set. First, we observe that a more preferable set of choices for $\kappa$ and $\lambda$ are around $1$ and $10\gamma$ respectively. We could see that if $\lambda$ is too small, due to the presence of transaction costs, the overall performance is less satisfactory. This shows that ignoring the trade off between return and transaction costs could yield a under-performing strategy. On the other hand, if $\lambda$ is too large, then we could not capture the possible investment opportunities since we are forced to stay close to the current portfolio. Another parameter $\kappa$ controls the  extent of robustness, or equivalently the risk of the portfolio. We observe that a relatively small or large choice of $\kappa$ may not give the best performance. It is intuitive that a large $\kappa$ may lead to a deterioration of portfolio return but from the plot of maximum drawdown, it could reduce the associated risk of the strategy simultaneously. If $\kappa$ is small, since MSCI covers the 2008 financial crisis period which leads to a significant drop in the assets' returns, this does not give the best portfolio performance as well. The presence of robustness could lead to the superior performance over other existing OLPS strategies.

\begin{figure}[p]
\includegraphics[scale=0.7]{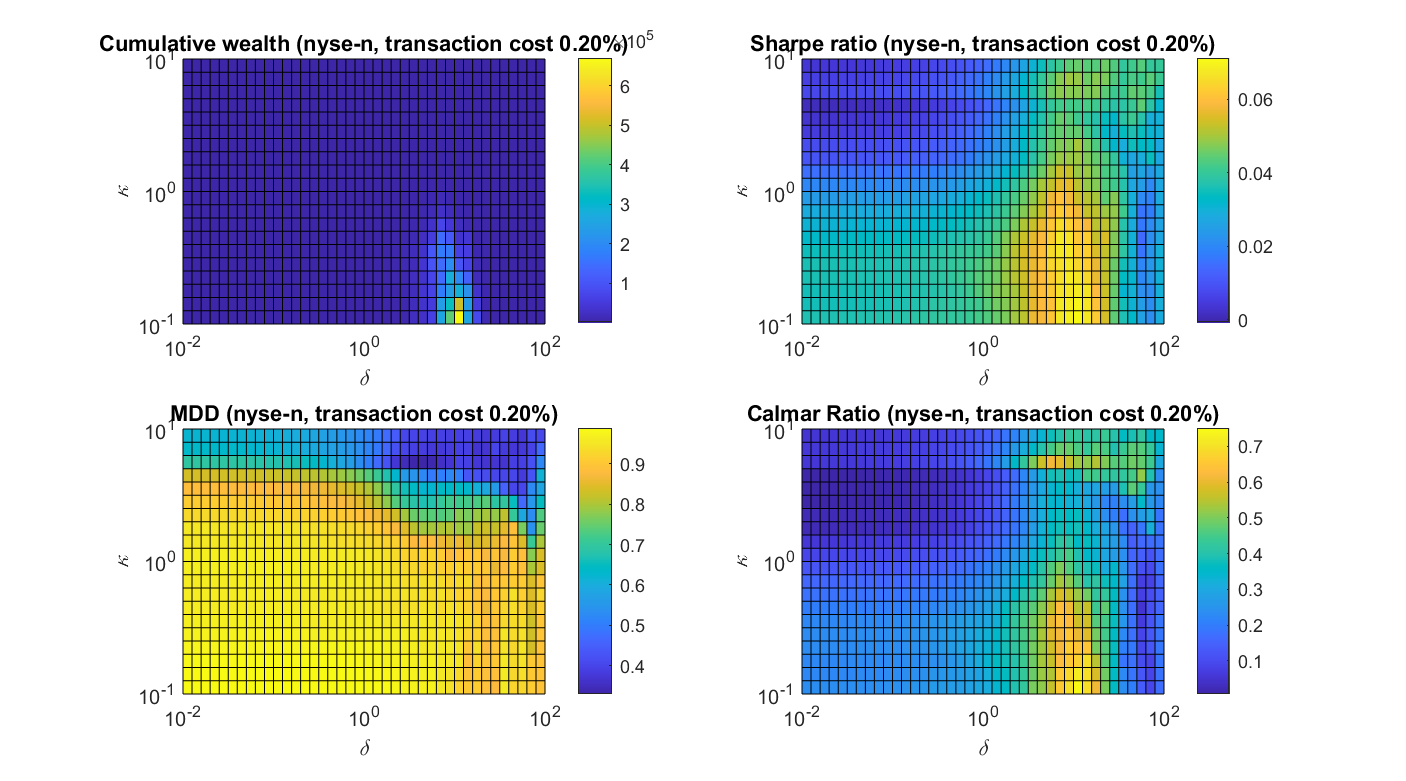}
\caption{Sensitivity on parameters on NYSE-N data set with $\gamma=0.2\%$} \label{fig:sensitivity_nyse-n_gamma_0.002}
\end{figure}

Figure~\ref{fig:sensitivity_nyse-n_gamma_0.002} shows the results for NYSE-N data set. First, we observe that a more preferable choice for $\lambda$ is still around $10\gamma$ but for $\kappa$, it becomes some values close to $0.1$. The observations on $\lambda$ is similar to the MSCI data set. That is, either too small or too large penalty on the transaction would yield a poor performance in general. The behaviour of $\kappa$ is slightly different from the MSCI data set. Note that it is still consistent that a large $\kappa$ would lead to a small cumulative wealth and maximum drawdown at the same time. However, a smaller choice of $\kappa$ is preferred. This could be explained by the fact that the NYSE-N data set covers a relatively long period and the assets experiences a general upward increasing trend. As we could observe that the magnitude of the cumulative return is in the order of $10^5$. Therefore, instead of a more robust strategy, a less robust and risk-averse strategy could be more preferable. 

\begin{figure}[p]
\includegraphics[scale=0.7]{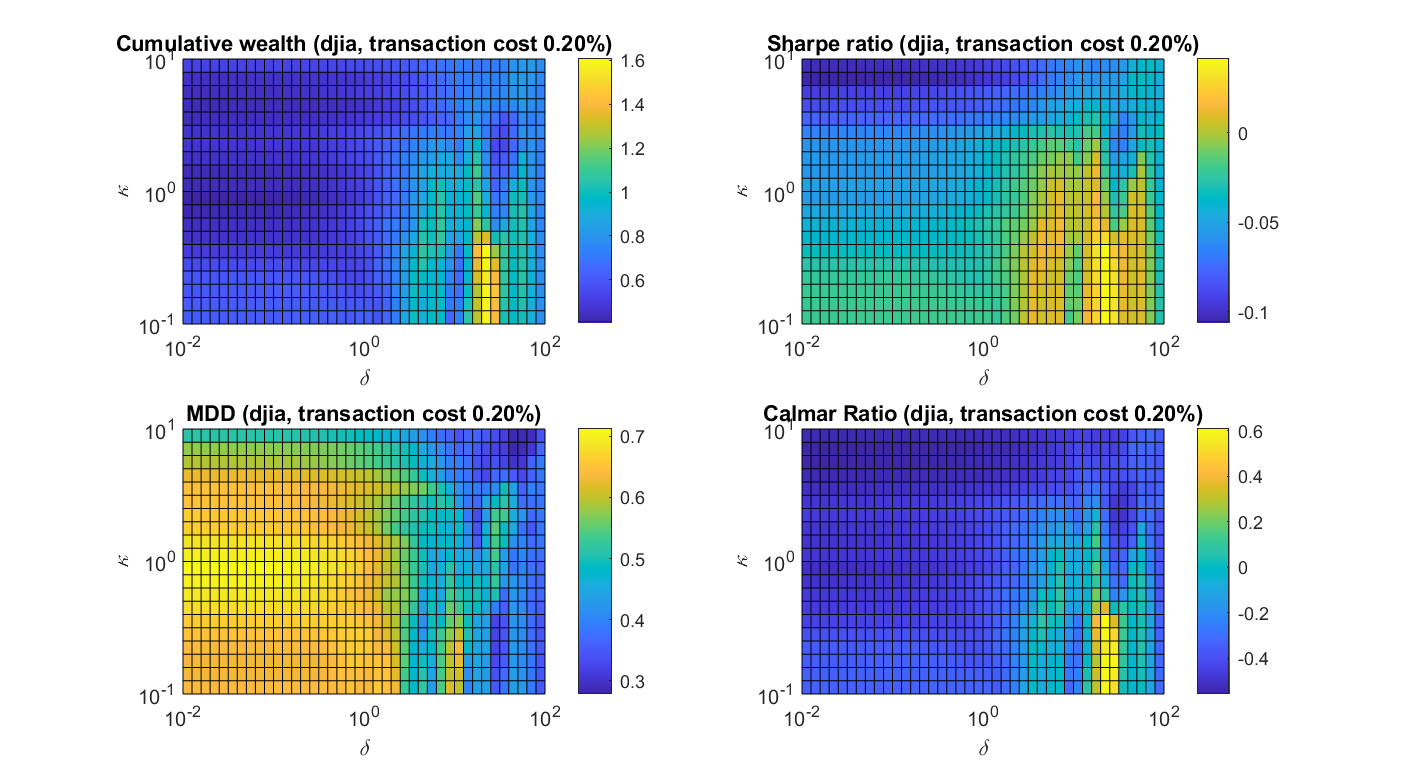}
\caption{Sensitivity on parameters on DJIA data set with $\gamma=0.2\%$} \label{fig:sensitivity_djia_gamma_0.002}
\end{figure}

\begin{figure}[p]
\includegraphics[scale=0.7]{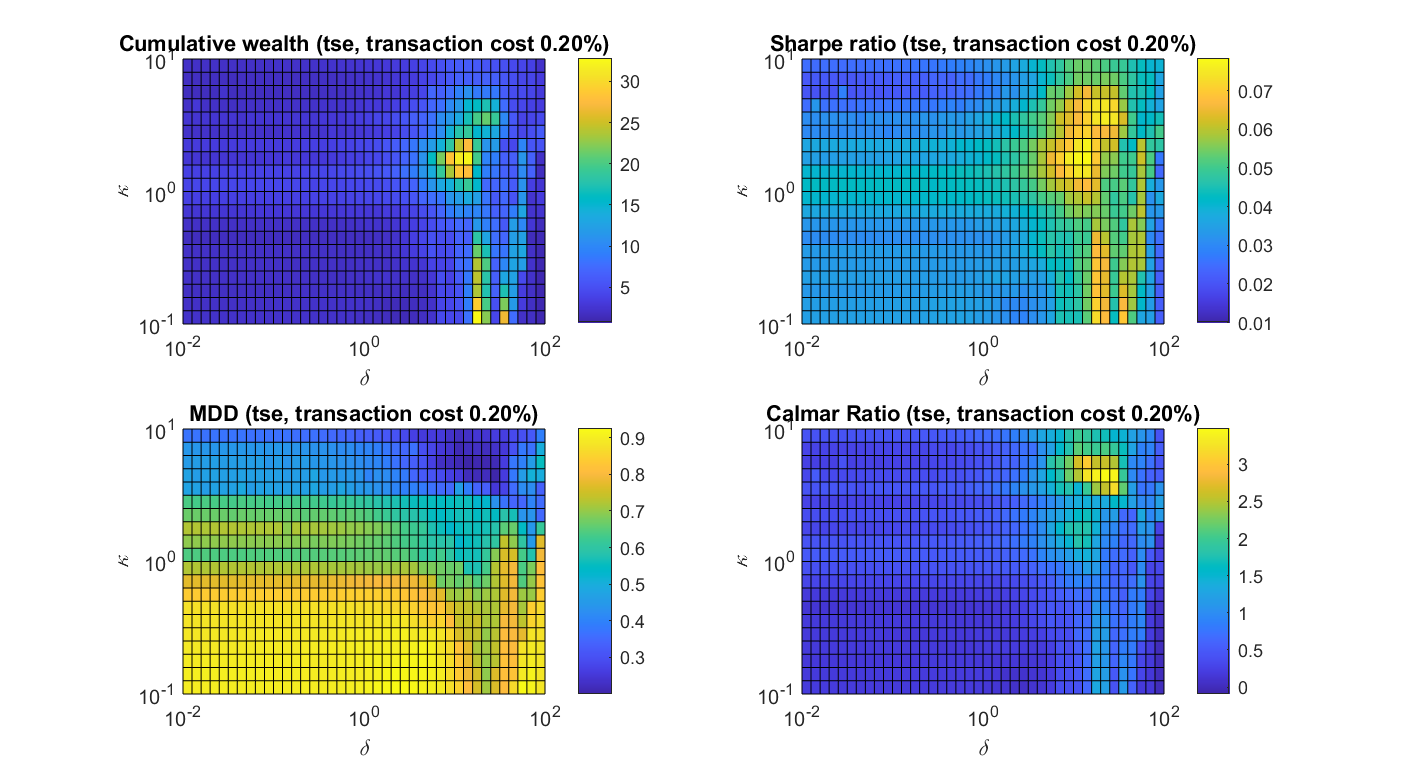}
\caption{Sensitivity on parameters on TSE data set with $\gamma=0.2\%$} \label{fig:sensitivity_tse_gamma_0.002}
\end{figure}

\begin{figure}[p]
\includegraphics[scale=0.7]{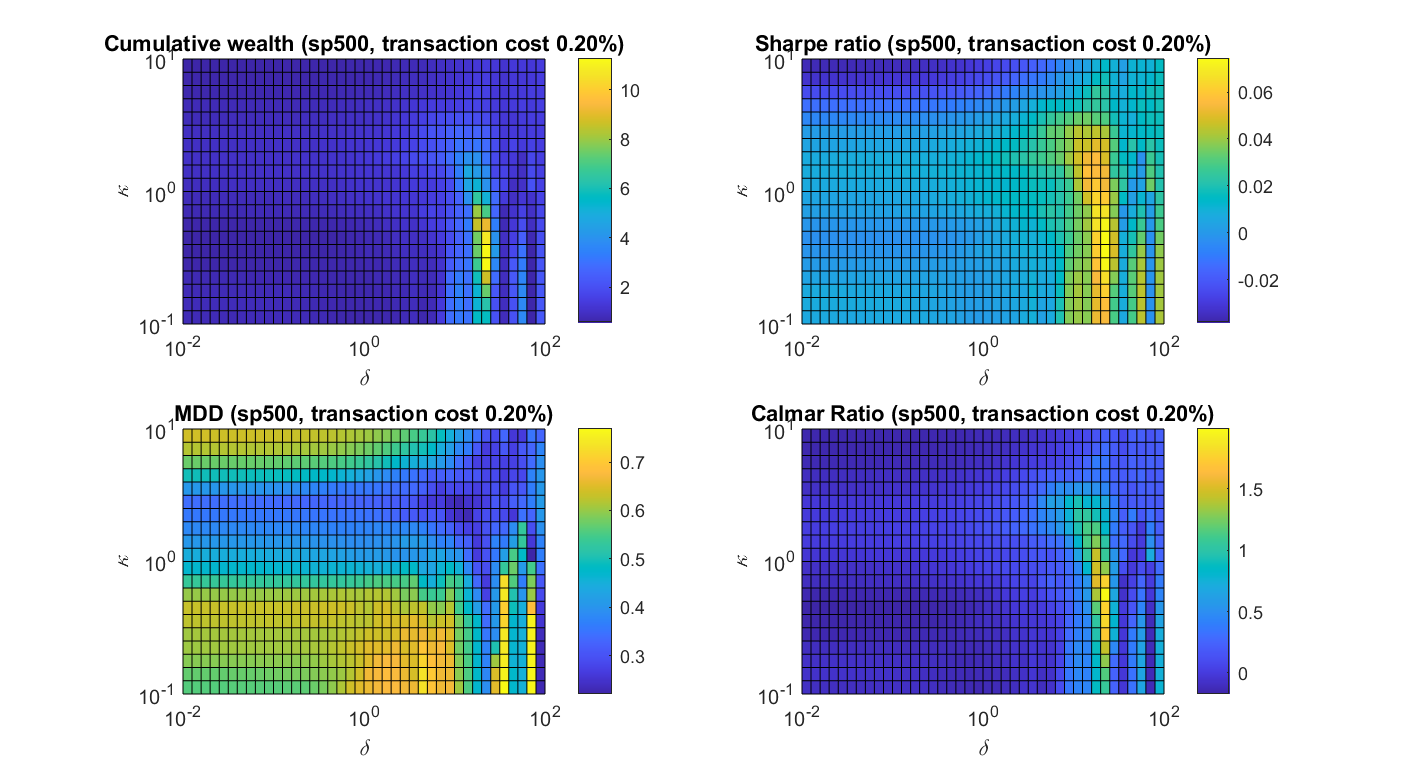}
\caption{Sensitivity on parameters on SP500 data set with $\gamma=0.2\%$} \label{fig:sensitivity_sp500_gamma_0.002}
\end{figure}

\begin{figure}[p]
\includegraphics[scale=0.7]{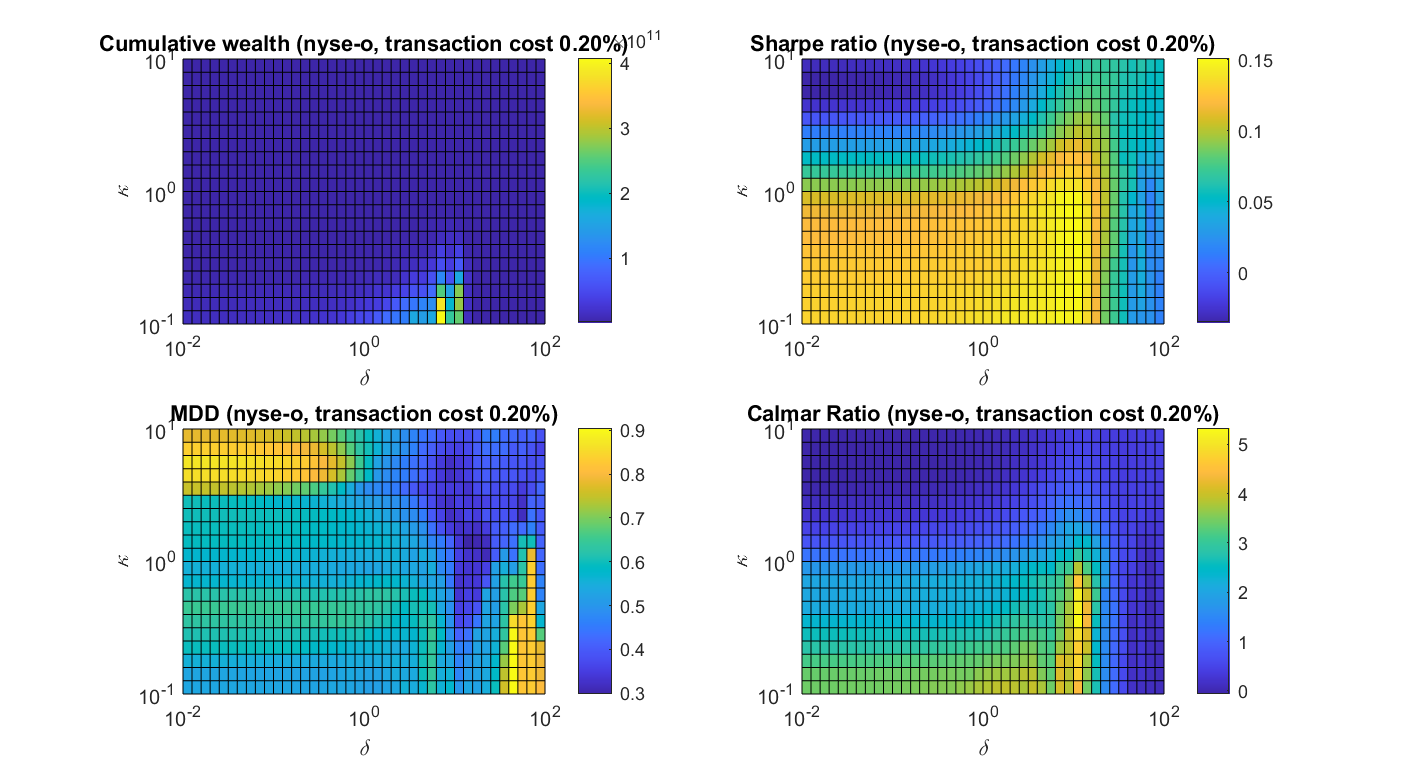}
\caption{Sensitivity on parameters on NYSE-o data set with $\gamma=0.2\%$} \label{fig:sensitivity_nyse-o_gamma_0.002}
\end{figure}

Figures~\ref{fig:sensitivity_djia_gamma_0.002}--\ref{fig:sensitivity_nyse-o_gamma_0.002} show the results for DJIA, TSE, SP500 and NYSE-O data sets respectively. Generally, we observe that if $\lambda=\delta\gamma$ is too small, due to the presence of transaction cost, the cumulative wealth may be poorer. Also, if $\lambda$ is too large, the optimal portfolio could not deviate from the current portfolio by a significant amount. Therefore, it may not be able to produce portfolios which are indeed profitable most of the time. We observe that the choice of $\lambda$ close to $10\gamma$ is more preferable in the benchmark data sets. However, the choice of $\kappa$, which corresponds to the extent of robustness, could be different across different data sets. For instance, in the DJIA data set, a smaller choice of $\kappa$ may yield a better portfolio performance. This may be due to reason that the stock prices in DJIA fluctuate and a less robust portfolio decision could capture the most recent market movement easily to obtain a satisfactory return. Similar to the discussion on NYSE-N, a smaller choice of $\kappa$ is preferable. On the other hand, for both TSE and SP500 data sets, there is market downturn during the selected period and a larger choice of $\kappa$ could yield more robust portfolios that hedge against unfavorable scenarios.

Note that depending on different data sets, the choice of parameters could have significant effect on the strategy performance. This is true, in general, for any OLPS strategies. Therefore, the choice of parameters could be critical to its performance and a default choice of parameter may not necessarily give a promising performance. This motivates us to investigate different adaptive schemes for choosing the parameters.

\subsection{The benchmark data sets} \label{appdx:benchmark_data_set}

\subsubsection{Other performance metric for Section~\ref{subsubsec:compare_algorithms}} \label{appdx:compare_algorithms}
\begin{figure}[t]
\centering
\begin{subfigure}{0.475\textwidth}
\centering
\includegraphics[scale=0.57]{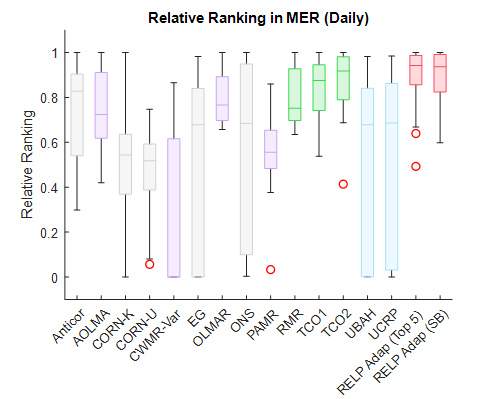}
\caption{MER (daily)} \label{fig:benchmark_mer_summary}
\end{subfigure}
\hfill
\begin{subfigure}{0.475\textwidth}
\centering
\includegraphics[scale=0.57]{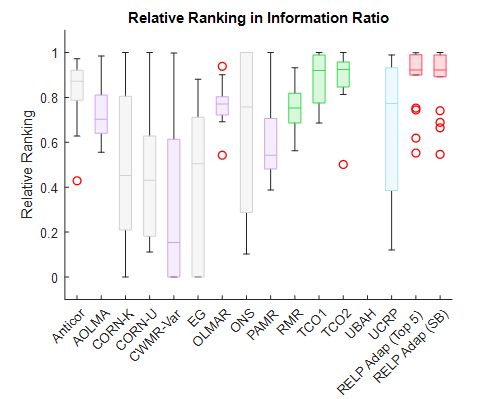}
\caption{Relative rankings on information ratio} \label{fig:benchmark_information_summary}
\end{subfigure}
\caption{Relative rankings on MER and information ratio} \label{fig:benchmark_mer_information_summary}
\end{figure}

We provide the results under the experimental setting in Section~\ref{subsubsec:compare_algorithms} on other performance metrics. To examine the average performance in return during the whole investment horizon, we can consider MER using the market strategy as the baseline strategy. Figure~\ref{fig:benchmark_mer_summary} summarizes the results of the MER for different OLPS strategies. We can still observe that our proposed strategies rank close to the top most of the time and it outperforms most of the existing OLPS strategies. The patterns are similar for other OLPS strategies to what we observed in the cumulative wealth.

We also compare the information ratio, which takes into account MER normalized by the standard deviation of market excess return. Hence, it could be viewed as the risk-adjusted MER. Since we are comparing the market strategy itself, the information ratio is not defined for UBAH. Figure~\ref{fig:benchmark_information_summary} gives the results for information ratio. Again, we observe that our proposed strategy could outperform many existing strategies in general with TCO2 being a close competitor. This demonstrates that our strategy could significantly outperform the market strategy to a large extent relative to other OLPS strategies.

\begin{figure}[t]
\centering
\includegraphics[scale=0.57]{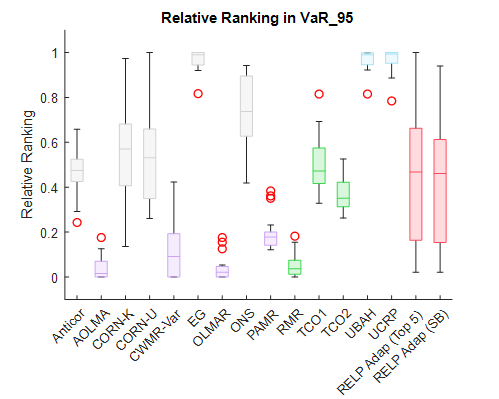}
\caption{$95\%$ value-at-risk} \label{fig:benchmark_var95_summary}
\end{figure}

\begin{figure}[t]
\hspace{-15mm}
\includegraphics[scale=0.6]{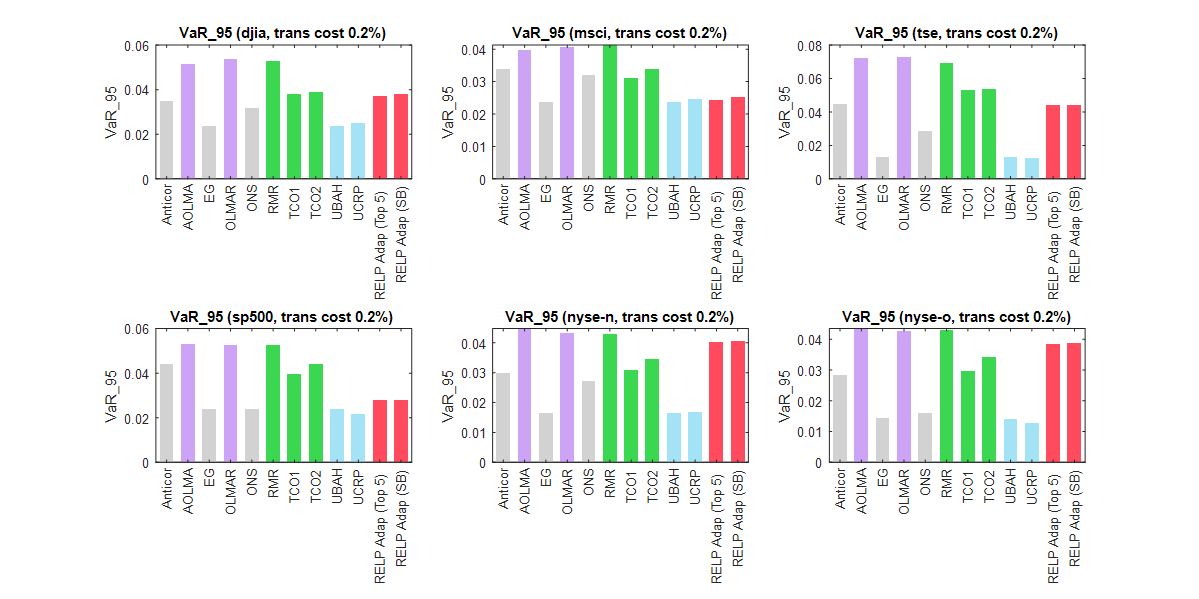}
\caption{$95\%$ value-at-risk of different strategies for the benchmark data sets with $\gamma=0.2\%$} \label{fig:benchmark_var95_gamma_2}
\end{figure}

Finally, Figure~\ref{fig:benchmark_var95_summary} presents the $95\%$ value-at-risk of the portfolio of different OLPS strategies. We observe that the two reference strategies usually have the best  (i.e., smallest) value-at-risk. Meanwhile, EG and ONS that include an regularization term also has a relatively small value-at-risk in general. This is seldom reported in the existing literature. For our proposed strategy, the value-at-risk appears to have a diverse performance among different settings. We typically observe that the $95\%$ value-at-risk is larger than other strategies in the setting with $\gamma=0$. This may be due to the reason that $\lambda=10\gamma=0$, which leads to more dramatic changes in portfolio weights. However, we remark that this is not a practical situation. Indeed, in the presence of transaction cost, the value-at-risk could have a similar or even lower value then many existing strategies. Figure~\ref{fig:benchmark_var95_gamma_2} shows the value-at-risk of selected strategies (with better cumulative wealth as observed in Figure~\ref{fig:benchmark_wealth_summary}) using the six benchmark data sets with $\gamma=0.2\%$. Except for NYSE-N and NYSE-O which involve a long investment horizon and yield returns with very large magnitudes (in the order of $10^8$ and $10^{6}$ respectively), we could observe that the our value-at-risk is in a low to medium regime. We also note that our strategies outperform TCO strategies, as a potential competitor, in many data sets.

\subsubsection{Full adaptive scheme} \label{appdx:full_adaptive_benchmark}
In this section, we present the additional results of the proposed OLPS strategy with adaptive schemes on both $\kappa$ and $\lambda$ when $\gamma=0.5\%$. Tables~\ref{table:benchmark_cumulative_wealth_gamma_5} to \ref{table:benchmark_calmar_gamma_5} provide the cumulative wealth, Sharpe ratio and Calmar ratio for different data sets under $\gamma=0.5\%$. First, from Table~\ref{table:benchmark_cumulative_wealth_gamma_5}, we observe that an increase in transaction cost substantially reduce the return from OLPS strategies that do not take transaction costs into account (e.g., OLMAR and RMR). We still observe that our proposed OLPS strategies could yield higher cumulative wealth than many existing strategies in most data sets. Table~\ref{table:benchmark_sharpe_gamma_5} also demonstrates our advantages in terms of Sharpe ratio. We observe that TCO1 with the default choice of parameter could perform well in general for these six benchmark data sets. However, as we have shown in Section~\ref{subsec:new_data} (and also in the next section), this is not true when we use our newly collected data sets. Finally, we observe that the Calmar ratio (Table~\ref{table:benchmark_calmar_gamma_5}) could perform well in most data sets but not as many as we have seen when $\gamma=0.2\%$. This may due to the fact that when the transaction cost increases, our adaptive scheme is designed to change the whole portfolio. This leads to a larger transaction costs when compared with $\gamma=0.2\%$. One way to remedy such a situation, as mentioned in the discussions from Section~\ref{sec:numerics}, is to use a sparser choice of parameters (e.g., only a few number of choices for $\kappa$ and $\lambda$ that capture more distinctive expert portfolio performances). Nevertheless, our proposed adaptive scheme allows the investor to input a range of parameters instead of a single one, which could lead to a more robust performance in data sets with distinctive features.

\begin{table}[t]\small
\centering
\ra{0.6}
\caption{Cumulative wealth comparison on six benchmark data sets when $\gamma=0.5\%$} \label{table:benchmark_cumulative_wealth_gamma_5}
\begin{tabular}{@{}rrrrrrr@{}} \toprule
Strategy   ($\gamma=0.5\%$) & DJIA                           & MSCI                           & TSE                            & SP500                          & NYSE-N                           & NYSE-O                           \\ \cmidrule{2-7}
EG                          & \cellcolor[HTML]{D3E4F4}0.7570 & \cellcolor[HTML]{E1EDF8}0.8989 & \cellcolor[HTML]{ECF4FB}1.5918 & \cellcolor[HTML]{E0ECF7}1.3136 & \cellcolor[HTML]{F8FBFD}1.73E+01 & \cellcolor[HTML]{FFFFFF}1.38E+01 \\
OLMAR                       & \cellcolor[HTML]{FFFFFF}0.0833 & \cellcolor[HTML]{FFFFFF}0.0135 & \cellcolor[HTML]{FFFFFF}0.0218 & \cellcolor[HTML]{FFFFFF}0.0045 & \cellcolor[HTML]{FFFFFF}1.60E-09 & \cellcolor[HTML]{FFFFFF}2.39E+00 \\
ONS                         & \cellcolor[HTML]{BDD7EE}1.0792 & \cellcolor[HTML]{EBF3FA}0.5980 & \cellcolor[HTML]{F5F9FD}0.8996 & \cellcolor[HTML]{D2E4F4}1.9150 & \cellcolor[HTML]{FDFEFF}5.85E+00 & \cellcolor[HTML]{FFFFFF}2.43E+01 \\
RMR                         & \cellcolor[HTML]{FFFFFF}0.0912 & \cellcolor[HTML]{FFFFFF}0.0110 & \cellcolor[HTML]{FFFFFF}0.0554 & \cellcolor[HTML]{FFFFFF}0.0017 & \cellcolor[HTML]{FFFFFF}2.21E-10 & \cellcolor[HTML]{FFFFFF}1.29E+00 \\
TCO1                        & \cellcolor[HTML]{C2DAF0}1.0104 & \cellcolor[HTML]{D9E8F6}1.1265 & \cellcolor[HTML]{F5F9FD}0.9076 & \cellcolor[HTML]{F3F8FC}0.5262 & \cellcolor[HTML]{BDD7EE}1.43E+02 & \cellcolor[HTML]{F6F9FD}2.33E+06 \\
TCO2                        & \cellcolor[HTML]{C9DFF2}0.8999 & \cellcolor[HTML]{E6F0F9}0.7575 & \cellcolor[HTML]{BDD7EE}5.3695 & \cellcolor[HTML]{EBF3FA}0.8723 & \cellcolor[HTML]{E8F1FA}5.02E+01 & \cellcolor[HTML]{FFFFFF}2.21E+04 \\
UBAH                        & \cellcolor[HTML]{D3E4F4}0.7606 & \cellcolor[HTML]{E1EDF8}0.9018 & \cellcolor[HTML]{ECF4FA}1.6049 & \cellcolor[HTML]{E0ECF7}1.3350 & \cellcolor[HTML]{F7FAFD}1.80E+01 & \cellcolor[HTML]{FFFFFF}1.44E+01 \\
UCRP                        & \cellcolor[HTML]{D1E4F4}0.7803 & \cellcolor[HTML]{E1EDF8}0.8834 & \cellcolor[HTML]{EEF5FB}1.4519 & \cellcolor[HTML]{DCEAF6}1.4865 & \cellcolor[HTML]{F6FAFD}2.13E+01 & \cellcolor[HTML]{FFFFFF}1.94E+01 \\
RELP-Adap-1                 & \cellcolor[HTML]{DCEAF6}0.6201 & \cellcolor[HTML]{C0D9EF}1.8461 & \cellcolor[HTML]{F4F9FD}0.9399 & \cellcolor[HTML]{DFECF7}1.3525 & \cellcolor[HTML]{C0D9EF}1.39E+02 & \cellcolor[HTML]{BDD7EE}1.54E+07 \\
RELP-Adap-2                 & \cellcolor[HTML]{C3DBF0}1.0001 & \cellcolor[HTML]{BDD7EE}1.9134 & \cellcolor[HTML]{F0F6FC}1.2435 & \cellcolor[HTML]{BDD7EE}2.7568 & \cellcolor[HTML]{D5E6F5}9.28E+01 & \cellcolor[HTML]{D0E3F3}1.11E+07 \\
\bottomrule
\end{tabular}
\end{table}
\begin{table}[t]\small
\centering
\ra{0.6}
\caption{Sharpe ratio comparison on six benchmark data sets when $\gamma=0.5\%$} \label{table:benchmark_sharpe_gamma_5}
\begin{tabular}{@{}rrrrrrr@{}} \toprule
Strategy   ($\gamma=0.5\%$) & DJIA                            & MSCI                            & TSE                             & SP500                           & NYSE-N                          & NYSE-O                         \\ \cmidrule{2-7}
EG                          & \cellcolor[HTML]{D3E5F4}-0.0384 & \cellcolor[HTML]{CCE0F2}-0.0089 & \cellcolor[HTML]{CCE1F2}0.0301  & \cellcolor[HTML]{C8DEF1}0.0115  & \cellcolor[HTML]{C0D9EF}0.0313  & \cellcolor[HTML]{EEF5FB}0.0373 \\
OLMAR                       & \cellcolor[HTML]{FFFFFF}-0.1375 & \cellcolor[HTML]{FCFEFF}-0.1591 & \cellcolor[HTML]{FFFFFF}-0.0254 & \cellcolor[HTML]{F7FAFD}-0.1110 & \cellcolor[HTML]{FAFCFE}-0.0738 & \cellcolor[HTML]{FDFEFF}0.0174 \\
ONS                         & \cellcolor[HTML]{BDD7EE}0.0102  & \cellcolor[HTML]{D1E3F3}-0.0240 & \cellcolor[HTML]{E9F2FA}-0.0014 & \cellcolor[HTML]{C0D9EF}0.0310  & \cellcolor[HTML]{C7DDF1}0.0177  & \cellcolor[HTML]{EBF3FA}0.0418 \\
RMR                         & \cellcolor[HTML]{FCFEFF}-0.1307 & \cellcolor[HTML]{FFFFFF}-0.1684 & \cellcolor[HTML]{F5F9FD}-0.0140 & \cellcolor[HTML]{FFFFFF}-0.1328 & \cellcolor[HTML]{FFFFFF}-0.0829 & \cellcolor[HTML]{FFFFFF}0.0145 \\
TCO1                        & \cellcolor[HTML]{BFD9EF}0.0065  & \cellcolor[HTML]{C7DDF1}0.0054  & \cellcolor[HTML]{D7E7F5}0.0181  & \cellcolor[HTML]{D0E3F3}-0.0104 & \cellcolor[HTML]{BDD7EE}0.0352  & \cellcolor[HTML]{BDD7EE}0.1013 \\
TCO2                        & \cellcolor[HTML]{C3DBF0}-0.0029 & \cellcolor[HTML]{CCE0F2}-0.0092 & \cellcolor[HTML]{BDD7EE}0.0464  & \cellcolor[HTML]{CADFF2}0.0053  & \cellcolor[HTML]{C0D9EF}0.0299  & \cellcolor[HTML]{D6E6F5}0.0696 \\
UBAH                        & \cellcolor[HTML]{D3E4F4}-0.0377 & \cellcolor[HTML]{CCE0F2}-0.0087 & \cellcolor[HTML]{CCE0F2}0.0310  & \cellcolor[HTML]{C7DDF1}0.0123  & \cellcolor[HTML]{BFD9EF}0.0318  & \cellcolor[HTML]{EDF4FB}0.0385 \\
UCRP                        & \cellcolor[HTML]{D0E3F3}-0.0322 & \cellcolor[HTML]{CCE0F2}-0.0094 & \cellcolor[HTML]{D5E5F4}0.0213  & \cellcolor[HTML]{C5DCF0}0.0181  & \cellcolor[HTML]{BFD8EF}0.0326  & \cellcolor[HTML]{E6F0F9}0.0479 \\
RELP-Adap-1                 & \cellcolor[HTML]{CEE2F3}-0.0277 & \cellcolor[HTML]{BED8EF}0.0342  & \cellcolor[HTML]{DCEAF6}0.0137  & \cellcolor[HTML]{C5DCF1}0.0174  & \cellcolor[HTML]{BED8EF}0.0347  & \cellcolor[HTML]{C0D9EF}0.0984 \\
RELP-Adap-2                 & \cellcolor[HTML]{BED8EF}0.0091  & \cellcolor[HTML]{BDD7EE}0.0356  & \cellcolor[HTML]{D6E7F5}0.0195  & \cellcolor[HTML]{BDD7EE}0.0375  & \cellcolor[HTML]{BFD8EF}0.0329  & \cellcolor[HTML]{C1D9EF}0.0973 \\
\bottomrule
\end{tabular}
\end{table}
\begin{table}[t]\small
\centering
\ra{0.6}
\caption{Calmar ratio comparison on six benchmark data sets when $\gamma=0.5\%$} \label{table:benchmark_calmar_gamma_5}
\begin{tabular}{@{}rrrrrrr@{}} \toprule
Strategy   ($\gamma=0.5\%$) & DJIA                            & MSCI                            & TSE                             & SP500                           & NYSE-N                          & NYSE-O                         \\ \cmidrule{2-7}
EG                          & \cellcolor[HTML]{DFECF7}-0.3343 & \cellcolor[HTML]{DAE9F6}-0.0392 & \cellcolor[HTML]{C9DFF2}0.3210  & \cellcolor[HTML]{D6E6F5}0.1183  & \cellcolor[HTML]{BED8EF}0.2219  & \cellcolor[HTML]{F5F9FD}0.2921 \\
OLMAR                       & \cellcolor[HTML]{FFFFFF}-0.7657 & \cellcolor[HTML]{FFFFFF}-0.6554 & \cellcolor[HTML]{FFFFFF}-0.5406 & \cellcolor[HTML]{FDFEFF}-0.6593 & \cellcolor[HTML]{FDFEFF}-0.5478 & \cellcolor[HTML]{FEFFFF}0.0399 \\
ONS                         & \cellcolor[HTML]{BDD7EE}0.1022  & \cellcolor[HTML]{E1EDF8}-0.1616 & \cellcolor[HTML]{E0ECF7}-0.0415 & \cellcolor[HTML]{C2DAF0}0.5282  & \cellcolor[HTML]{CADFF2}0.0774  & \cellcolor[HTML]{EDF4FB}0.5061 \\
RMR                         & \cellcolor[HTML]{FEFFFF}-0.7476 & \cellcolor[HTML]{FFFFFF}-0.6709 & \cellcolor[HTML]{FAFCFE}-0.4471 & \cellcolor[HTML]{FFFFFF}-0.7171 & \cellcolor[HTML]{FFFFFF}-0.5816 & \cellcolor[HTML]{FFFFFF}0.0116 \\
TCO1                        & \cellcolor[HTML]{C4DCF0}0.0114  & \cellcolor[HTML]{D4E5F4}0.0542  & \cellcolor[HTML]{DFECF7}-0.0236 & \cellcolor[HTML]{E4EFF8}-0.1541 & \cellcolor[HTML]{BDD7EE}0.2309  & \cellcolor[HTML]{BDD7EE}1.7940 \\
TCO2                        & \cellcolor[HTML]{CEE2F3}-0.1206 & \cellcolor[HTML]{DDEBF7}-0.1010 & \cellcolor[HTML]{BDD7EE}0.5062  & \cellcolor[HTML]{DEEBF7}-0.0418 & \cellcolor[HTML]{C2DAF0}0.1759  & \cellcolor[HTML]{E3EEF8}0.7794 \\
UBAH                        & \cellcolor[HTML]{DEEBF7}-0.3300 & \cellcolor[HTML]{DAE9F6}-0.0381 & \cellcolor[HTML]{C9DEF1}0.3280  & \cellcolor[HTML]{D6E6F5}0.1280  & \cellcolor[HTML]{BED8EF}0.2239  & \cellcolor[HTML]{F5F9FD}0.3036 \\
UCRP                        & \cellcolor[HTML]{DCEAF6}-0.2958 & \cellcolor[HTML]{DAE9F6}-0.0454 & \cellcolor[HTML]{CFE2F3}0.2268  & \cellcolor[HTML]{D0E3F3}0.2446  & \cellcolor[HTML]{C0D9EF}0.1942  & \cellcolor[HTML]{F2F8FC}0.3675 \\
RELP-Adap-1                 & \cellcolor[HTML]{E0ECF7}-0.3509 & \cellcolor[HTML]{BFD8EF}0.4137  & \cellcolor[HTML]{DFECF7}-0.0174 & \cellcolor[HTML]{D8E7F5}0.0880  & \cellcolor[HTML]{BED8EF}0.2213  & \cellcolor[HTML]{C1DAEF}1.6959 \\
RELP-Adap-2                 & \cellcolor[HTML]{C5DCF0}0.0001  & \cellcolor[HTML]{BDD7EE}0.4314  & \cellcolor[HTML]{DAE9F6}0.0611  & \cellcolor[HTML]{BDD7EE}0.6123  & \cellcolor[HTML]{C0D9EF}0.2055  & \cellcolor[HTML]{C3DBF0}1.6486 \\
\bottomrule
\end{tabular}
\end{table}

\subsection{The new data sets} \label{appdx:new_data_set}

In this section, we provide the results on the new data sets when the transaction cost $\gamma$ is $0.5\%$. Table~\ref{table:new_data_comparison_gamma_5} shows the cumulative wealth, Sharpe ratio and information ratio from several OLPS strategies. Note that the results are similar to what we have seen in Section~\ref{subsec:new_data} when $\gamma=0.2\%$. We observe that the cumulative wealth from our proposed algorithm performs generally better than the existing strategies. As explained in Section~\ref{subsec:new_data}, our adaptive scheme may result in a larger standard deviation since we keep track of the experts with the best cumulative wealth and the expert could be risk-taking. This happens when there is some asset outperforming the others substantially. Therefore, this makes sense that the Sharpe ratio may not be very large. Finally, information ratio, as a measure of the out-performance against the market strategy, of our strategy is generally better than other OLPS strategies (in SP500-21 and HSI). This demonstrates that our strategy could beat the market strategy in many cases. Similar to the observations with $\gamma=0.2\%$, the market strategy outperforms all the OLPS strategies and RELP-Adap-1 could maintain an overall cumulative wealth close to the market strategy. 
\begin{table}[t]\small
\centering
\ra{0.6}
\caption{Comparison of new data sets on cumulative wealth (CW), Sharpe ratio (SR) and information ratio (IR) when $\gamma=0.5\%$} \label{table:new_data_comparison_gamma_5}
\begin{tabular}{@{}rrrr|rrr|rrr@{}} \toprule
Strategy         & \multicolumn{3}{c}{SP500-21}                                                                       & \multicolumn{3}{c}{HSI}                                                                             & \multicolumn{3}{c}{CSI}                                                                            \\
($\gamma=0.5\%$) & CW                             & SR                              & IR                              & CW                              & SR                              & IR                              & CW                             & SR                              & IR                              \\ \cmidrule{2-10}
EG               & \cellcolor[HTML]{DCEAF6}3.7502 & \cellcolor[HTML]{BDD7EE}0.0666  & \cellcolor[HTML]{BED8EF}0.0338  & \cellcolor[HTML]{D5E6F5}8.3030  & \cellcolor[HTML]{BDD7EE}0.0514  & \cellcolor[HTML]{BDD7EE}0.0422  & \cellcolor[HTML]{BED8EF}2.7066 & \cellcolor[HTML]{BED8EF}0.0259  & \cellcolor[HTML]{C0D9EF}-0.0114 \\
OLMAR            & \cellcolor[HTML]{FFFFFF}0.0010 & \cellcolor[HTML]{FAFCFE}-0.1734 & \cellcolor[HTML]{F9FCFE}-0.2502 & \cellcolor[HTML]{FFFFFF}0.0000  & \cellcolor[HTML]{FCFDFF}-0.2158 & \cellcolor[HTML]{FEFEFF}-0.2997 & \cellcolor[HTML]{FFFFFF}0.0000 & \cellcolor[HTML]{F9FCFE}-0.2289 & \cellcolor[HTML]{F8FBFE}-0.3090 \\
ONS              & \cellcolor[HTML]{F2F8FC}1.4083 & \cellcolor[HTML]{CBE0F2}0.0134  & \cellcolor[HTML]{CFE2F3}-0.0480 & \cellcolor[HTML]{FBFDFE}0.7977  & \cellcolor[HTML]{CBE0F2}-0.0073 & \cellcolor[HTML]{D1E3F3}-0.0610 & \cellcolor[HTML]{D8E7F5}1.6359 & \cellcolor[HTML]{C1DAEF}0.0106  & \cellcolor[HTML]{C3DBF0}-0.0303 \\
RMR              & \cellcolor[HTML]{FFFFFF}0.0005 & \cellcolor[HTML]{FFFFFF}-0.1961 & \cellcolor[HTML]{FFFFFF}-0.2817 & \cellcolor[HTML]{FFFFFF}0.0000  & \cellcolor[HTML]{FFFFFF}-0.2308 & \cellcolor[HTML]{FFFFFF}-0.3102 & \cellcolor[HTML]{FFFFFF}0.0000 & \cellcolor[HTML]{FFFFFF}-0.2560 & \cellcolor[HTML]{FFFFFF}-0.3464 \\
TCO1             & \cellcolor[HTML]{E9F2FA}2.3799 & \cellcolor[HTML]{C6DDF1}0.0324  & \cellcolor[HTML]{C7DDF1}-0.0124 & \cellcolor[HTML]{EFF5FB}3.2406  & \cellcolor[HTML]{C3DBF0}0.0275  & \cellcolor[HTML]{CADFF2}-0.0256 & \cellcolor[HTML]{C5DCF1}2.3943 & \cellcolor[HTML]{BFD8EF}0.0210  & \cellcolor[HTML]{BDD7EE}0.0006  \\
TCO2             & \cellcolor[HTML]{DAE9F6}4.0423 & \cellcolor[HTML]{C2DAF0}0.0479  & \cellcolor[HTML]{C2DAF0}0.0135  & \cellcolor[HTML]{F0F6FC}2.9871  & \cellcolor[HTML]{C4DBF0}0.0249  & \cellcolor[HTML]{C9DFF2}-0.0201 & \cellcolor[HTML]{E0ECF7}1.3174 & \cellcolor[HTML]{C2DAF0}0.0080  & \cellcolor[HTML]{C1DAEF}-0.0172 \\
UBAH             & \cellcolor[HTML]{DDEAF7}3.7149 & \cellcolor[HTML]{BED8EF}0.0666  & \cellcolor[HTML]{FFFFFF}NaN     & \cellcolor[HTML]{D7E7F5}7.8918  & \cellcolor[HTML]{BED8EF}0.0512  & NaN                             & \cellcolor[HTML]{BDD7EE}2.7242 & \cellcolor[HTML]{BDD7EE}0.0261  & NaN                             \\
UCRP             & \cellcolor[HTML]{E4EFF8}2.9941 & \cellcolor[HTML]{BFD9EF}0.0589  & \cellcolor[HTML]{D0E3F3}-0.0537 & \cellcolor[HTML]{F3F8FC}2.4731  & \cellcolor[HTML]{C4DBF0}0.0254  & \cellcolor[HTML]{D0E3F3}-0.0587 & \cellcolor[HTML]{C5DCF0}2.4120 & \cellcolor[HTML]{BED8EF}0.0229  & \cellcolor[HTML]{C2DAF0}-0.0221 \\
RELP-Adap-1      & \cellcolor[HTML]{CEE1F3}5.3079 & \cellcolor[HTML]{C2DAF0}0.0503  & \cellcolor[HTML]{BFD9EF}0.0253  & \cellcolor[HTML]{BDD7EE}12.9185 & \cellcolor[HTML]{BED8EF}0.0472  & \cellcolor[HTML]{C1DAEF}0.0227  & \cellcolor[HTML]{C4DCF0}2.4485 & \cellcolor[HTML]{C1DAEF}0.0109  & \cellcolor[HTML]{BED8EF}-0.0041 \\
RELP-Adap-2      & \cellcolor[HTML]{BDD7EE}7.0658 & \cellcolor[HTML]{BFD9EF}0.0587  & \cellcolor[HTML]{BDD7EE}0.0343  & \cellcolor[HTML]{DBE9F6}7.2150  & \cellcolor[HTML]{BED8EF}0.0486  & \cellcolor[HTML]{C4DCF0}0.0068  & \cellcolor[HTML]{DAE8F6}1.5679 & \cellcolor[HTML]{C7DDF1}-0.0141 & \cellcolor[HTML]{C1DAEF}-0.0173  \\
\bottomrule
\end{tabular}
\end{table}

\clearpage
\bibliographystyle{elsarticle-harv}
\linespread{1}

\bibliography{references}

\end{document}